\documentclass[times,12pt]{article}
\usepackage{amsmath,amssymb,amsfonts,latexsym,amsthm,enumerate,url}
\usepackage{graphicx}
\usepackage{xcolor}
\usepackage{hyperref}
\usepackage{float}
\usepackage{tablefootnote}

\newcommand{\beq}[1]{\begin{equation}\label{#1}}
\newcommand{\enq}[0]{\end{equation}}

\newcommand{\remove}[1]{}

\newcommand{\comment}[1]{}

\begin{document}

\title {Questions and Concerns About Google's Quantum Supremacy Claim}

\author {Gil Kalai, Yosef Rinott, and Tomer Shoham}
\maketitle

\begin{abstract}

In October 2019, {\it Nature} published a paper \cite {Aru+19} describing 
an experimental work that was performed at Google.
The paper claims to demonstrate quantum (computational) supremacy on a 53-qubit quantum computer. Since then we 
have been involved in a long-term project to study various statistical
aspects of the Google experiment. In \cite {RSK22} we studied Google’s statistical framework
that we found to be very sound and offered some technical improvements. This document describes three main concerns (based on statistical analysis) about the Google 2019 experiment. The first concern is that  
 the data do not agree with Google's noise model (or any other specific model). The second concern is that a crucial simple formula for a priori estimation of the fidelity  
 seems to involve an unexpected independence assumption, and yet it gives very accurate predictions. 
 The third concern is about 
 statistical properties of the calibration process.\footnote {Research supported by ERC grant 834735. We thank  
Carsten Voelkmann for many corrections and helpful suggestions.}  
\end{abstract}

\newpage

\section {Introduction}
 
The 2019 paper ``Quantum supremacy using a programmable
superconducting processor'' \cite {Aru+19} claimed that
Google's  Sycamore quantum computer performed a certain computation in about 200 seconds, while
a state-of-the-art classical supercomputer
would take, according to the Google team's estimates,
approximately 10,000 years to perform the same computation. 
Google's Sycamore quantum computer performed a \textit{sampling task}; that is,
it generates random bitstrings of length 53
with considerable noise,
from a certain discrete probability distribution supported on all such $2^{53}$ bitstrings. 
(A bitstring is a vector of zeroes and ones.)
Google's specific sampling task is referred to as random circuit sampling (RCS, for short).
In 2021, 
a team from USTC repeated the
Google 2019 experiment with its {\it Zuchongzhi} superconducting processor    
\cite {Wu+21,Zhu+22}, and claimed to achieve
an even stronger form of quantum advantage compared to the Google experiment.

In the Google experiment the Sycamore quantum computer produced samples, 
where each sample consists 
of several hundred thousand binary strings, for each of around 1000 random circuits. For each sample, a certain statistic  called the {\it linear cross-entropy fidelity estimator} denoted by ${\mathcal F}_{XEB}$, was computed. Based on the value of this estimator it was concluded that the samples produced by the quantum computer represent ``quantum supremacy."    
As for Google's quantum supremacy claim, several groups \cite{PanZha21,PCZ22,ZSW20,KPY21,KPZY21,PGN+19,Gao+21}  
have since produced classical algorithms that are ten 
orders of magnitude faster than those used in the Google paper. 
This was achieved, for example, by Pan, Chen, and Zhang \cite {PCZ22} in 2021.

\subsubsection* {Putting Google's quantum supremacy claim under the microscope}

This paper presents statistical analysis that may shed light  
on the quality and reliability of the data and the statistical methods 
of the Google experiment. 
The first author (Kalai) has been raising some concerns about the Google experiment since September 2019 (when \cite {Aru+19} and \cite {Aru+19S} were leaked) and a few months after the publication of the Google paper we initiated what has become a long-term project to study various statistical aspects of the Google experiment and to scrutinize the Google paper. This is a good place to mention that Google's quantum supremacy claim appeared to refute Kalai's theory regarding quantum computation (\cite {Kal18,Kal20,Kal22}) and Kalai's specific prediction that NISQ systems cannot demonstrate ``quantum supremacy.''  This fact influenced and may have biased Kalai's assessment of Google's quantum supremacy claim. (Recent improved classical algorithms have largely refuted Google's quantum supremacy claim and therefore the Google results no longer refute Kalai's theory.)

Here is a brief review of our previous papers. In our first paper \cite {RSK22} we mainly studied fidelity estimators and made a preliminary comparison of the empirical distribution of the samples and that of the Google noise model.  
Our recent paper \cite {KRS22d} provided a detailed description of data and information regarding the Google experiment and listed some confirmations, refutations, weaknesses, and concerns.

This paper studies three central concerns. In Section 2 we provide a brief background on the Google experiment and some of its statistical ingredients. In Section 3 we describe our first concern: the data does not agree with Google's noise model (or any other specific model). In Section 4 we describe our second concern regarding the striking predictive power of a certain crucial a priori estimate (``Formula (77)") for the ${\mathcal F}_{XEB}$ fidelity estimator.
In Section 5 we discuss our third concern regarding the remarkable effectiveness and other statistical properties of the 2-gate calibration.

\subsubsection *{A few problems of general interest}

Scrutinizing a scientific work  necessarily involves 
punctiliousness and nitpicking, but there are several issues that we consider to be of general interest. 

(a) What is the statistical methodology for analyzing
samples obtained from noisy quantum computers 
and for finding 
appropriate models to describe the empirical data?

(b) We suggest that the statistical independence assumption in a certain predictive model might be unrealistic. What could be the scientific framework and methodology to study this matter?

(c) We find it surprising that a local optimization process (namely, a process that separately optimizes each variable) of a function of many variables, reaches a critical point. 
What could be further tools to study this matter?

(d) What are the tools to study whether an empirical behavior is non-stationary and perhaps even inherently unpredictable?


(e) What are the appropriate methodology and ethics for scrutinizing 
major scientific works, and how is it possible to bridge the gap between theoreticians (like us) and experimentalists? 

\newpage

\section {Google's quantum supremacy claim}
\subsection {A brief background}
\label {s:g1}
We will now give a brief summary of the Google experiment,
Google's noise model,
Google's ${\mathcal F}_{XEB}$ linear cross-entropy fidelity estimator,
and Google's Formula (77) in \cite {Aru+19S} for predicting the fidelity
of a circuit from the fidelity
of its components. For more details the reader is referred to \cite {RSK22,Aru+19,Aru+19S}.

The Google experiment is based on the building of a quantum computer (circuit), 
with $n$ superconducting qubits,
that performs $m$ rounds of computation.
The computation is carried out by 1-qubit and 2-qubit gates. 
At the end of the computation the qubits are measured, leading 
to a string of zeroes and ones of length $n$. 
The ultimate experiment was for $n=53$ and $m=20$. It involved
1113 1-qubit gates and 430 2-qubit gates. For that experiment  
the Google team produced a sample of three million 0-1 vectors of length 53. 

\begin{figure}[h]
\centering
\includegraphics[scale=0.5]{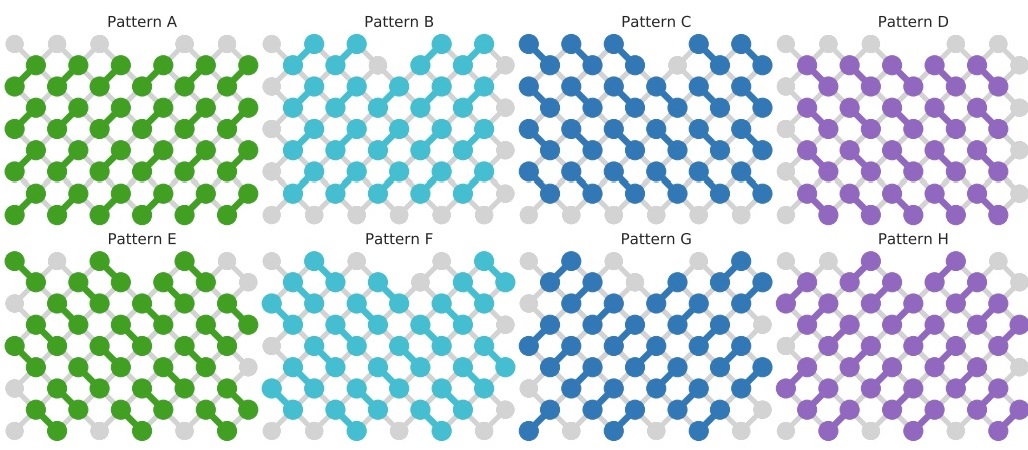}
\caption{Figure S25 from \cite {Aru+19S}. The different patterns of a single 2-gate layer for $n=53$.}
\label{S25}
\end{figure}

The circuits used in the Google experiment had the following structure. 
The qubits were arranged on a planar grid, and so a single qubit was identified via two 
coordinates, like qubit $(3,3)$. The circuits had two types of layers:
one type of layer consists of 2-gates acting on pairs of (neighboring) qubits. After each such layer of 2-gates there was another layer of randomly chosen 1-gates acting on every qubit. (The number $m$ of layers of 2-gates is referred to as the depth of the circuit.) The layers of 1-gates consist of the ``programmable" ingredient in the experiment.\footnote { 
The entire randomness in the Google experiment for more than 900 circuits is determined by the choice of random 1-gates for ten large random circuits. The same random 1-gates used for the $k$th experiment ($k=0,1,\dots,9$) of the largest full circuits 
with $n=53$ qubits, and depth  $m=20$, were also used for all types of circuits (i.e., full, elided, and patch circuits), patterns, number of qubits $n$, and depth $m$.} The layers of 2-gates are fixed throughout the experiment according to a certain pattern. The pattern {\bf EFGH} was used for 
circuits with numbers of qubits between 12 and 53 and of depth $m=14$, 
and a new pattern {\bf ABCDCDAB} was later used for $n=53$ qubits and depths $m=12,14,16,18,$ and $20$. 
Each letter (like {\bf E}) corresponds to a fixed set of 2-gates acting in parallel on the qubits; see Figure
\ref {S25}. (For more details, see \cite {KRS22d}.) 
As part of their experiment the Google team also studied two simplified versions of their  circuits, ``elided circuits" and ``patch circuits" that were based on removing some of the 2-gates of the ``full"  circuits.  In particular, for patch circuits the qubits were divided into two separate non-interacting patches.

Every circuit $C$ with $n$ qubits 
describes a probability distribution ${\bf P}_C(x)$ for 0-1 vectors of length $n$. 
(In fact, it describes a $2^n$-dimensional vector of complex amplitudes; for every 0-1 vector
$x$, there is an associated amplitude $z(x)$ and ${\bf P}_C(x)=|z(x)|^2.$)
The quantum computer enables one to sample according to the probability distribution ${\bf P}_C(x)$,
with a considerable amount of noise. When $n$ and $m$ are not too large, classical simulations enable one to compute the amplitudes themselves (and hence the probabilities ${\bf P}_C(x)$). Google's quantum supremacy claim is based on the fact that these classical simulations quickly become infeasible as $n$ and $m$ grow. 

When $C$ is a random circuit, the probability distribution ${\bf P}_C(x)$ behaves like a Porter--Thomas distribution; namely, the individual probabilities behave as if they have random statistically independent values drawn from the exponential distribution (that are then normalized).
Of course, an instance of a Porter--Thomas distribution depends on $2^n$ random variables, 
while the $2^n$ values of ${\bf P}_C(x)$ depend on the circuit $C$ whose description requires a polynomial number of parameters in $n$. 

The Google basic noise model is 
\begin {equation}
\label{e:gnm}
{\bf N}_C(x) =  \phi {\bf P}_C(x) + (1-\phi) 2^{-n},
\end {equation}
where $\phi$ is the {\it fidelity}, a parameter that roughly describes the quality of the sample. Roughly speaking, Google's noise model assumes that if all components of the quantum computer operate without errors then a bitstring will be drawn according to ${\bf P}_C(x)$, and if an error occurs then the quantum computer will produce a random uniformly distributed bitstring. 

{\bf Remark:} For a noisy quantum circuit, if  
$\rho$ is the density matrix of the ideal state (the state of a noiseless quantum computer running the circuit $C$) and $  \psi$ is the density matrix for the noisy state, 
then the fidelity $\phi$ is defined by \begin {equation}
\label {e:fid}
\phi =   \langle \psi \left | \rho \right | \psi \rangle.
\end {equation}
We note that the fidelity $\phi$ itself cannot be read from the distribution of the samples produced by the quantum computer. Given an unlimited sample size (or samples large enough such that the empirical distribution is a good approximation to the noisy distribution) and unlimited computational power,  
it is an interesting question to find the best way to estimate the fidelity. 
Another important question that is very relevant to NISQ experiments is to find the best ways to estimate the fidelity given sample sizes that are only polynomial (in ($1/\phi$)).

Based on their noise model (and the fact that the distribution ${\bf P}_C$ is an instance of a Porter--Thomas distribution), the Google paper describes a statistic called the {\it linear cross-entropy estimator}, 
denoted by ${\cal F}_{XEB}$ (and sometimes simply by $XEB$).
Once the quantum computer produces a sequence ${\bf \tilde x}$ of $N$ 
bitstrings ${\bf \tilde x}= ({\bf \tilde x}^{(1)},{\bf \tilde x}^{(2)},\dots, {\bf \tilde x}^{(N)})$, the estimator ${\cal F}_{XEB}$ of the fidelity is computed as follows:

\begin {equation}
\label {e:fxeb}
{\cal F}_{XEB}({\bf \tilde x})=\frac{1}{N}\sum _{i=1}^N 2^n{\bf P}_C({\bf \tilde x}^{(i)}) -1.
\end {equation}
Computing ${\cal F}_{XEB}$ requires knowledge of ${\bf P}_C(x)$ for sampled bitstrings. 

The Google supremacy claim is also based on the following a priori prediction of the fidelity of a circuit based on the probabilities of error for the individual components (Formula (77) in 
\cite {Aru+19S}):

\begin {equation}
\label {e:77}
~\hat \phi~=~ \prod_{g \in {\cal G}_1} (1-e_g) \prod_{g \in {\cal G}_2} (1-e_g) \prod_{q \in {\cal Q}} (1-e_q).
\end {equation}

Here ${\cal G}_1$ is the set of 1-gates, ${\cal G}_2$ is 
the set of 2-gates, and
${\cal Q}$ is the set of qubits. For a gate $g$, the term $e_g$ in the 
formula refers to the probability of an error 
of the individual gate $g$. 
For a qubit $q$, $e_q$ is the probability of a readout error when we measure the qubit $q$.

The Google supremacy paper \cite {Aru+19} made two crucial claims regarding the ultimate 53-qubit samples.

\begin {itemize}
\item[(A)] The fidelity $\phi$ of their sample is above $1/1000$. 
 

\item [(B)] Producing a sample of similar fidelity 
would take 10,000 years on a classical supercomputer.
\end {itemize}
For claim (A) regarding the value of $\phi$, 
the argument  relies on an 
extrapolation argument that has two ingredients. 
One ingredient is a few hundred experiments in the classically tractable regime, namely, the 
regime where the  probability distribution ${\bf P}_C$ can be
computed by a classical computer and the performance of the 
quantum computer can be tested directly. The other
ingredient is the theoretical formula (\ref {e:77}) for predicting the fidelity.
According to the paper, the fidelity of 
entire circuits closely agrees with the prediction of Formula (\ref {e:77}) 
(Formula (77) in \cite {Aru+19S}) 
with a deviation below 10--20 percent. 
There are more than 700 reported experiments in the
classically tractable regime, including ones carried out on simplified
circuits (which are easier to simulate on classical computers). These experiments
support the claim that the
prediction given by Formula (77) for the fidelity is indeed very robust and applies
to the 53-qubit circuit in the supremacy regime.

For claim (B) regarding the classical difficulty, the Google team
mainly relies on extrapolation from the running time of a
specific algorithm they used. They also rely on the computational complexity 
support for the assertion
that the task at hand is asymptotically difficult.

{\bf Remarks:} 1. Claim (B) and with it the entire quantum supremacy claim, have largely been refuted by the research of several groups \cite{PanZha21,PCZ22,ZSW20,KPY21,KPZY21,PGN+19,Gao+21} (among others) 
that exhibited classical algorithms that are ten 
orders of magnitude faster than those used in the Google paper. 
This was achieved, for example, by Pan, Chen, and Zhang \cite {PCZ22} in 2021.
See also Section \ref {s:gao}.

2. Google's ${\cal F}_{XEB}$ fidelity estimator given by \eqref {e:fxeb} which is used for the full and elided circuits does not apply for the patch circuits. For patch circuits the ${\cal F}_{XEB}$ fidelity estimator is used separately for each patch and the two estimates are multiplied. 
\subsection {The distribution and the size-biased distribution}
\label {s:dsbd}
It is useful to distinguish between
\begin {itemize}
\item [(i)] the Porter--Thomas distribution ${\bf P}_C(x)$ on a space of size $2^n$, 
\item [(ii)] the real distribution of probabilities ${\bf P}_C(x)$ over all possible $2^n$ bitstrings, and
\item
[(iii)] the size-biased distribution, which is the real distribution of the probabilities $({\bf P}_C(x))$ 
when $x$ is sampled according to the Google noise model \eqref{e:gnm}.
\end {itemize}

In the first item, for every bitstring $x$ we have a probability ${\bf P}_C(x)$. In the second item 
we consider only these $2^n$ probabilities as a random set of real numbers. The $2^n$ probabilities ${\bf P}_C(x)$ over all $2^n$ bitstrings behave like a normalized sample 
of size $2^n$ of an exponential random variable. 

The size-biased distribution based on the Google noise model \eqref {e:gnm} (item (iii))  
behaves 
as the real distribution
with density function

\begin {equation}
\label {e:size-biased}
\phi x e^{-x} + (1-\phi) e^{-x} .  
\end {equation}

The empirical size-biased distribution from the Google bitstrings agrees very well with 
\eqref {e:size-biased}
(see Figure \ref{fig:hists} below, Figure S32 in \cite {Aru+19S} and Figure 9 in \cite {RSK22}). As we explain in \cite {RSK22} the size-biased distribution is stable to major changes of the ``full" $2^n$ distribution. For example, if you only sample bitstrings where the number of ones is  a prime number this will not change the size-biased distribution but will entail a major modification of the original Porter--Thomas distribution, by making most of the probabilities equal to zero.  

\subsection {The work of Gao et al.}
\label {s:gao}
Gao, 
Kalinowski, Chou, Lukin, Barak, and Choi \cite {Gao+21} 
studied random circuit sampling and considered three 
quantities for noisy quantum circuits: the first quantity is the fidelity $\phi$ defined by Equation \ref {e:fid}, the second quantity is the linear cross-entropy estimator ${\cal F}_{XEB}$, and the third quantity is the probability of no errors  denoted by $  p_{no~err}$. A basic observation of Gao et al. is that when you apply depolarizing noise to the gates, the resulting distribution has a positive correlation to the ideal distribution (and hence this leads to a positive ${\cal F}_{XEB}$ value). Applying such depolarization noise on a set of 2-gates that would split the circuit into two parts leads to a sort of ``patch" circuit for which one can make the computation separately on every patch, and this provides quick classical algorithms to create samples with high values of ${\cal F}_{XEB}$. The paper thus shows that there are ``adversarial" methods that allow us to achieve samples with high values of ${\cal F}_{XEB}$ 
even without having the ability to compute the individual amplitudes. 

The paper \cite {Gao+21} describes various additional reasons why the effect of gate errors leads to positive correlations with the ideal distribution, and in general leads to strict inequalities:
\begin {equation} 
\label {e:gao+}
{\cal F}_{XEB} > \phi > p_{no~err}.
\end {equation} 

{\bf Remarks:} 1.  Referring to their algorithm for quickly producing samples with high values of ${\cal F}_{XEB}$, Gao et al. wrote:  ``Remarkably, the XEB value of our algorithm generally improves for larger quantum circuits, whereas that of noisy quantum devices quickly deteriorates. Such scaling continues to hold when the number of qubits is increased while the depth of the circuit and the error-per-gate are fixed." In other words, asymptotically, when $n$ is large the adversarial classical method of \cite {Gao+21} outperforms the quantum computer.

2. Regarding the comparison with current experiments, Gao et al. wrote: ``Specifically, we present an efficient classical algorithm that achieves high XEB values, namely 2--12\% of those obtained in the state-of-the-art experiments, within just a few seconds using a single GPU machine." For Google's ultimate supremacy experiment ($n=53$, $m=20$), Gao et al. (\cite {Gao+21}, Table I) produced in 0.6 seconds on a single GPU machine samples with an ${\cal F}_{XEB}$ fidelity of 0.000185 which is 8\% of Google's claimed fidelity of 0.00224.

3. It is possible that allowing their algorithm more running time may outperform Google's ${\cal F}_{XEB}$ value on available classical computers. We also note that Gao et al.'s method used ``patch circuits" and it may also be possible to base their algorithm on ``elided circuits" to achieve efficient algorithms for adversarial methods that outperform Google's fidelity claims.

\subsection {Some findings of this paper}

\begin {enumerate}
\item 
There is a large gap between the samples of the Google experiment and the Google noise mode, and any other specific noise model. (Section \ref {s:empirical}.) 
The gap between the empirical distribution and the model is asymmetric. (Section \ref {s:asym}.)
\item 
There are large fluctuations of the empirical behavior which are not understood. Consequently, there is evidence that the distance
between the Google noise model and uniform distribution is smaller (when
the number of qubits is $n > 16$) than the distance between the experimental
samples and the Google noise model. (Section \ref {s:t}, \ref {s:multi}.)
\item The empirical behavior of the samples is not stationary.
(Section \ref {s:chaos}.)
\item 
While the empirical distribution is not stationary the ${\cal F}_{XEB}$ fidelity is stable along the samples. Moreover, ``high energy events" that lead to abrupt increase in errors that are reported for later experiments with the Sycamore quantum computer cannot be detected in the 2019 quantum supremacy experiment. (Section \ref {s:chaos}.)

\item The predictive power of Formula (77) for the XEB fidelity estimates is statistically  surprising: the subsumed independence between components of 
systems such as quantum computers, 
which are known to be sensitive to noise and errors caused by interactions
with their environment, is striking. (Section \ref {s:77a}.)

\item The is systematic bias between the predictions of Formula (77) for patch circuits and the experimental data; the Google explanation for this bias is not convincing. (Sections \ref {s:patch1}, \ref {s:patch-cal}.)

\item The behavior of the fidelities of the two patches for patch circuits is very different; This appears to be in tension with Formula (77). (Section \ref {s:patch2}.)

\item The success of the experiments fully depends on the very large effect of the calibration adjustments. (Section \ref {s:1gc}.) There are large differences between the effects of the calibration adjustments for different 2-gates, and even for different appearances of the same 2-gate.

\item The calibration adjustments are surprisingly effective, especially given the stated local nature of the calibration. Mathematically speaking, we witness  a local optimization process reaching a critical point of a function depending on hundreds of parameters (Sections \ref {s:ue1}, \ref{s:2gc}.)

\end {enumerate}

\subsection {Brief suggestions for future quantum supremacy experiments}
For future experiments we propose that the following steps be carried out, with a clear separation between them.
\begin {itemize} 
\item [Step A:] Calibrate the circuits based on experiments on 1- and 2- qubit circuits.

\item [Step B:] Create a fully randomized collection of random circuits where for each circuit new random 1-gates are selected.
(This step can be carried out in a blind way as suggested in \cite {RSK22} and Section 5.1 of \cite {KRS22d}.) 

\item [Step C:] Run the quantum computer, produce the bitstrings and compute the fidelities.
\end {itemize}

In the 2019 quantum supremacy experiment, the same random 1-gates for ten circuits were applied to all circuits and were chosen prior to the calibration stage. A new calibration variant (``parallel XEB") was developed at the same time as the experiment and was completed a few days after the samples were produced (see Section 2.4 of \cite {KRS22d}).

We also propose: a) Produce large samples (10M--50M bitstrings) for a few of the circuits allowing to study the empirical distribution. 
b) Provide a detailed account of the fidelities of all components that are used in Formula (77). 
c) Provide the raw data for the calibration experiments. This would allow us partial checking of the the calibration process, without revealing the calibration process itself. (See Section 5.2 of \cite {KRS22d}.)

\newpage

\section {Concern I: The data do not fit the Google model (or any other specific noise model)}
\label {s:dist}
\subsection {Sampling computational problems}

The common definition of sampling problems is the following definition, taken from  Aaronson and Chen \cite {AarChe16} (see also, Lund et al. \cite {Lun+17}):


\begin {quotation}
``Problems where the goal is to sample an $n$-bit string, either
exactly or approximately, from a desired probability distribution." 
\end {quotation}

Bouland et al. \cite {Bou+19} described the framework for proposals for demonstrating quantum supremacy based on sampling computational tasks, as follows:

\begin {quotation}

``Proposals for quantum supremacy have a common framework. The computational task is to sample from the output distribution $D$ of some experimentally feasible quantum process or algorithm (on some given input). To establish quantum supremacy we must show hardness (that is, no efficient classical algorithm can sample from any distribution close to $D$) and verification (that is, an algorithm can check that the experimental device sampled from an output distribution close to $D$).” 

\end {quotation}

In this section we show that the random bitstrings produced in the Google quantum supremacy experiment do not obey these definitions.\footnote {Aaronson and Gunn \cite {AarGun19} and Gao et al. \cite {Gao+21} studied the extent to which 
quantum supremacy can be based on the computational hardness of sampling with a high value of ${\mathcal F}_{XEB}$, without referring to the question of whether a specific sampling task was achieved.}
The empirical distribution of the samples produced by the quantum computer is quite different from the Google noise model and from any other specific model we are aware of. We also study further interesting statistical properties of the samples provided by the Google experiment. 

\subsection {The empirical distribution for $n=12,14$}
\label {s:empirical}
\begin{figure}
	\begin{center}
		\includegraphics[width=\textwidth]{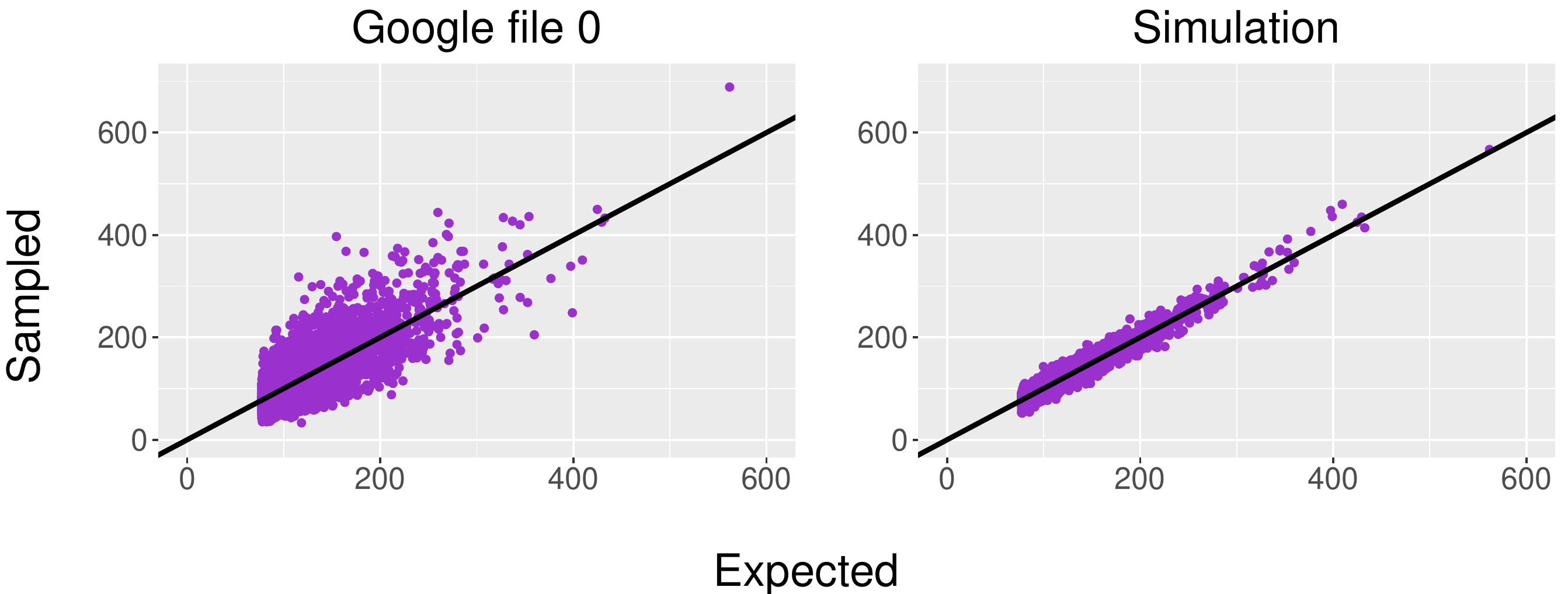}
		\caption{The left-hand side scatterplots display theoretical vs. empirical frequencies of the Google sample (file 0) with $n=12$. The right-hand side scatterplots display  theoretical vs. our simulated empirical frequencies according to Google's noise model \eqref{e:gnm} with $\phi=0.3701$.}
		\label{fig:scattt}
	\end{center}
\end{figure}

For an experimental circuit $C$, given a bitstring $x$, we denote by $O(x)$ the number of occurrences of $x$ in the sample. 
The {\it empirical distribution} is a discrete probability distribution on the set of bitstrings $x$, 
defined by $O(x)/N$, where $N$ is the size of the sample. 
Recall that $n$ denotes the number of qubits and $M=2^n$ denotes the number of bitstrings.   
In \cite {RSK22} we used Pearson's chi-square test statistic to compare 
the number of occurrences 
with the Google noise model. 
Pearson's chi-square test statistic is 
\begin {equation}
{\cal X}^2
=\sum_{i=1}^M
\frac  {(O(x_i)-N {\bf N}_C(x_i))^2}
{N {\bf N}_C(x_i)},
\end {equation}

\noindent
and its asymptotic distribution for fixed $M$ and large $N$ is chi-square with $M-1$ degrees of freedom (or $M-2$ if $\phi$ is estimated). (${\bf N}_C$ is defined by \eqref{e:gnm}.)
For $n=12$, the $\chi ^2$-value we obtained was around 40,000, where the expected value for samples according to ${\bf N}_C(x)$ is around 4,000; see Table \ref {T:distances-full}. 
The conclusion was 
that the Google model (\ref {e:gnm}) does not provide an adequate description of the data and that additional models should be explored.  
The large discrepancy between the empirical distribution and the Google noise model is evident also from Figure \ref {fig:scattt}. 

We considered a few other notions of distance to compare the number of occurrences with Google's noise model, and with other noise models,
$$ L_1 = \sum_{i=1}^M |N{\bf N}_C(x_i)-O(x_i)|, \qquad L_2 = \sqrt{\sum_{i=1}^M (N{\bf N}_C(x_i)-O(x_i))^2}, $$
$$ KL = \sum_{i=1}^M \frac{O(x_i)}{N} \text{log}\left( \frac{O(x_i)/N}{{\bf N}_C(x_i)} \right),$$ and the Pearson correlation. (KL refers to the Kullback–Leibler divergence.)
The outcomes are given in Table \ref {T:distances-full}.


In \cite {RSK22}, Section 6, we considered refined error models that take into account the readout errors. When we consider our best refined error model (the asymmetric readout noise model) the chi-square distance between the noise model and the empirical distribution is improved by a small amount from 41,190 and 54,286 for $n=12,14$ respectively to 36,786 and 50,093  (Table \ref {T:distances-full}). For $n=12,14$, the chi-square distance between samples of the uniform distribution and the Google noise model is only modestly larger than the distance between the empirical samples and the Google noise model, and we expect that for larger values of $n$ the quantum computer samples will be further away from the Google model (and any specific model we are aware of) than samples drawn from the uniform distribution (see remark in Section \ref{s:t}). It will be interesting to find a better match to the empirical distribution using detailed gate errors and this can be carried out with simulations of noisy circuits. 

\subsection {The estimator $T$ and the empirical variance}  
\label {s:t}
In \cite {RSK22} we considered another estimator $T$ for the fidelity based on the occurrence numbers 
(This estimate does not require knowledge of ${\bf P}_C(x_i)$.)
Since the expected variance of an 
$\exp (1)$ variable is 1, 
for fidelity level $\phi$ the expected variance of a distribution according to the Google noise model is $\phi^2 \cdot 2^{-2n}$. For a sample of size $N$, the expected variance of the sample will be larger than the variance of the distribution, but the variance of the distribution can be estimated from the variance of the sample. The unbiased (normalized) estimator $T^2$ for $\phi^2$ that takes into account the sampling size is defined as follows: 
\begin {equation}
\label {e:t}
T^2:=\frac{M(M+1)}{(N^2-N)(M-1)}\left(\sum_{i=1}^M O(x_i)^2 -N-(N^2-N)/M\right). 
\end {equation}

Under the Google noise model, $T^2$ is an unbiased estimator for $\phi^2$. 
The effect of readout and gate errors is expected to increase $T$ by a few percent. This is the case for the detailed readout models considered in Section 6 of \cite {RSK22} (Table \ref {tab:T_table}), and this is also the case for the contribution of gate errors under some reasonable statistical assumptions about them. We can estimate $T^2$ using \eqref {e:t} 
from the empirical data and this gives a pretty good estimate when $n \le 22$; 
this was carried out in \cite {RSK22} Section 4.7, and the results are that the values of $T$ for the Google bitstrings are much higher compared to the Google noise model. For example, as seen from Table \ref {tab:T_table} for $n=12, 18, 22$, $T$ is roughly 30\%, 50\% and 80\% higher compared to the Google model value. (This suggests that the variance of the actual noisy signal produced by the quantum computer is much larger than what the Google noise model gives, and, for example, for $n=22$, $T^2$ is more than three times larger compared to the ${\cal F}_{XEB}$ estimator.) This additional variance represents a substantial form of noise (that we denote below by ${\bf N}_T$) which is poorly understood. 

{\bf Remark:} If $T^2$ is indeed a good estimator for the variance it would follow that 
for values of $n$ which are not very small, the uniform distribution is ``closer" than the empirical distribution to the distribution described by the Google noise model.  When $n$ is larger than $14$ we can only estimate the $\ell_2$-distance under the (reasonable) assumption that the added noise ${\bf N}_T$ leading to the large variance is uncorrelated to ${\bf P}_C$, and the additional assumption that $T^2$ (normalized) estimates the variance. The expected value of the (normalized) $\ell_2$ distance between the uniform distribution and the distribution given by the Google noise model is $\phi$
(and it is concentrated near $\phi$ for a random circuit). Under our assumptions, the $\ell_2$ distance between the empirical distribution and the distribution described by the Google model is $\sqrt{T^2-\phi^2}$, which is larger than $\phi$ if $T> \sqrt 2\phi$. 

This suggests that for $n>14$ samples from the uniform distributions will be closer (and much closer as $n$ grows) to the Google noise model  compared to the experimental samples produced by the Sycamore quantum computer. (This conclusion applies not only to the Google noise model but to any specific noise model we are aware of.)\footnote {The work of Greg Valiant and Paul Valiant  \cite {ValVal17} that studied estimation of various statistical parameters like entropy, variance, and distances between the empirical distribution and a model, using samples of sublinear size, could be relevant here.}

\begin{table}[]
\centering
\resizebox{\textwidth}{!}{
 \begin{tabular}{||c| c| c| c| c| c| c| c||} 
 \hline
 $n$ & (77) & \multicolumn{2}{|c|}{Google file} & \multicolumn{2}{c|}{Simulation - Google model} & \multicolumn{2}{c|}{Simulation - Readout model} \\ \hline \hline 
        ~  & ~  &AVG XEB& AVG $T$ &AVG XEB& AVG $T$ &AVG XEB& AVG $T$ \\ [0.5ex] 
 \hline\hline
12&0.3862&0.3701&0.4689&0.3727&0.3712&0.3721&0.3742\\
14&0.3320&0.3298&0.4392&0.3316&0.3306&0.3315&0.3342\\
16&0.2828&0.2721&0.3917&0.2720&0.2722&0.2719&0.2755\\
18&0.2207&0.2442&0.3557&0.2442&0.2442&0.2442&0.2479\\
20&0.1875&0.2184&0.3210&0.2181&0.2179&0.2182&0.2216\\
22&0.1554&0.1650&0.2989&0.1651&0.1642&0.1649&0.1685\\
24&0.1256&0.1407&0.2838&0.1407&0.1312&0.1408&0.1264\\
 \hline
\end{tabular}}
\caption{Formula (77)'s prediction, average ${\cal F}_{XEB}$, and average $T$. Each simulation is based on 100 repetitions, 10 for each Google file. $\phi$ of the Google model in the simulations is taken to be the average ${\cal F}_{XEB}$ reported in the table, and $\phi_ro$ of the symmetric readout model is $\phi/(1-0.038)^n-\phi$ where $\phi$ is the average ${\cal F}_{XEB}$.}
\label{tab:T_table}
\end{table}



\subsection {The empirical size-biased distribution}

\begin{figure}[h]
	\begin{center}
		\includegraphics[width=13cm, height=7cm]{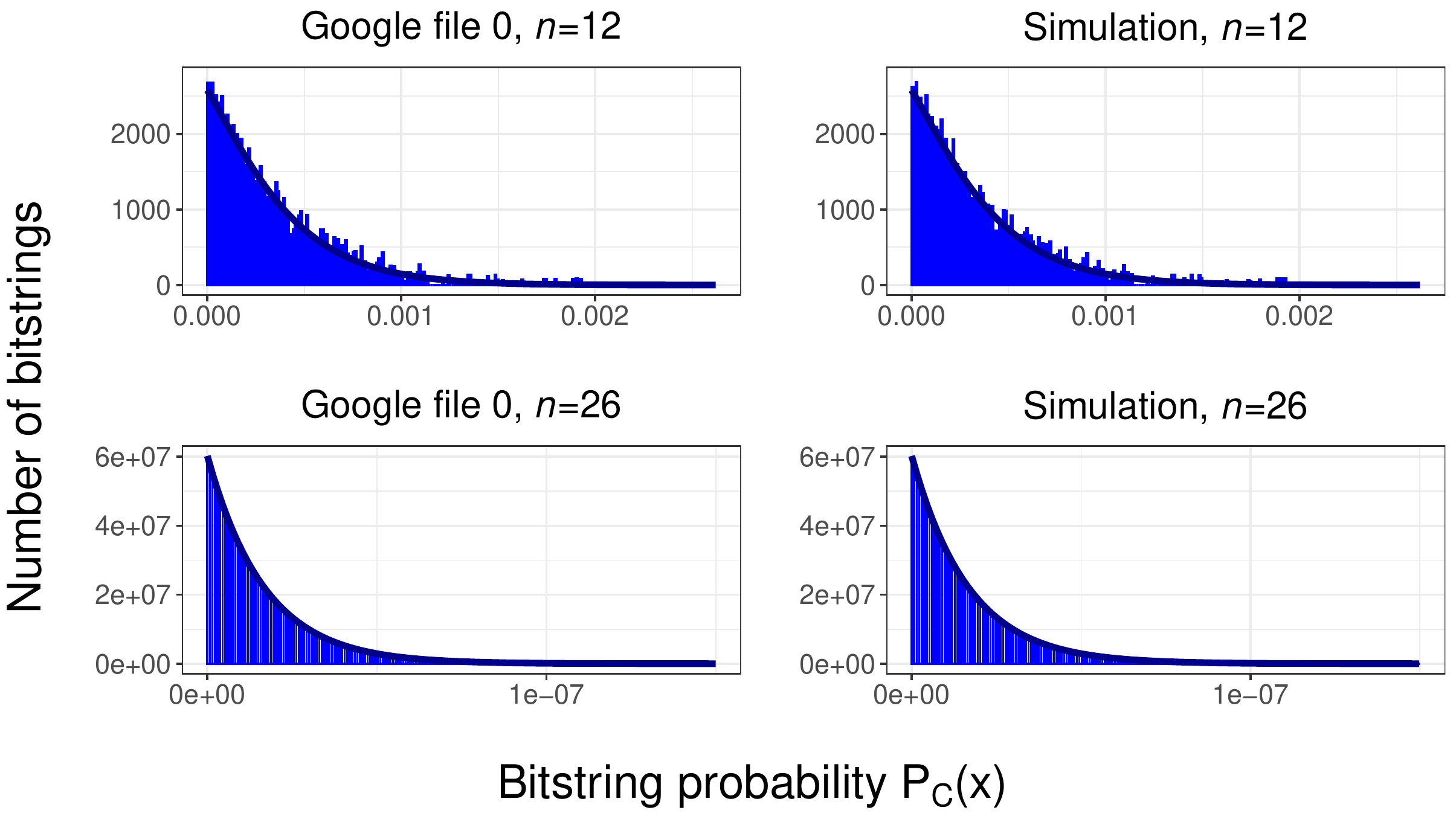}
		\caption{Histograms of $\{P_C({\bf \tilde x_i)}\}_{i=1}^N$ with $N=500,\!000$ from $\{ {\bf \tilde x_i}\}_{i=1}^N$ 
  of Google's file no. 0 $(n=12)$ (upper left) and $n=26$ (lower left). On the right is a sample generated according to Google's noise model  with $\phi=0.3701$ (upper right) and $\phi=0.1024$ (lower right). The number of cells is 200. The dark blue curves represent the asymptotic density of 
  the size-biased distribution.}
		\label{fig:hists}
	\end{center}
\end{figure}

The size-biased  distribution for a (noiseless) Porter--Thomas distribution is a real distribution with density function $xe^{-x}$. Under Google's noise model, when the fidelity level is $\phi$ the size-biased distribution is $\phi xe^{-x}+(1-\phi)e^{-x}$. (See Section \ref {s:dsbd}.) Given the sample ${\bf \tilde x^{(i)}}$, we refer to the real numbers 
\[
\{ {\bf P}_C({{\bf \tilde x}^{(i)}}) \},
\]

\noindent
as the empirical size-biased distribution.
While the empirical distribution of the Google samples is very different from the Google model, the coarse behavior of the empirical size-biased distribution  is very close to the model. (This is demonstrated in Figure \ref {fig:hists} below and in Figure S32 in \cite {Aru+19S}.) When $n$ is large, most bitstrings are occurring only once in the sample, and we cannot distinguish between the size-biased  empirical distributions coming from the Google files compared to simulation based on the Google model. 
For small values of $n$, we can exhibit the behavior of the Google data by considering the different behaviors in different scales, and the good agreement in the histogram (Figure \ref {fig:hists}) also reflects smoothing. 
We note also that when the fidelity is small, the empirical size-biased  distribution is dominated by samples from a uniform distribution and so the agreement of the empirical size-biased distribution with the theory does not convey much information other than that the noise is not correlated with the ``signal." 


\subsubsection {The behavior of the empirical distribution 
on different scales}
\label {s:multi}

\begin{figure}[h]
\centering
\includegraphics[width=\textwidth,height=\textheight,keepaspectratio]{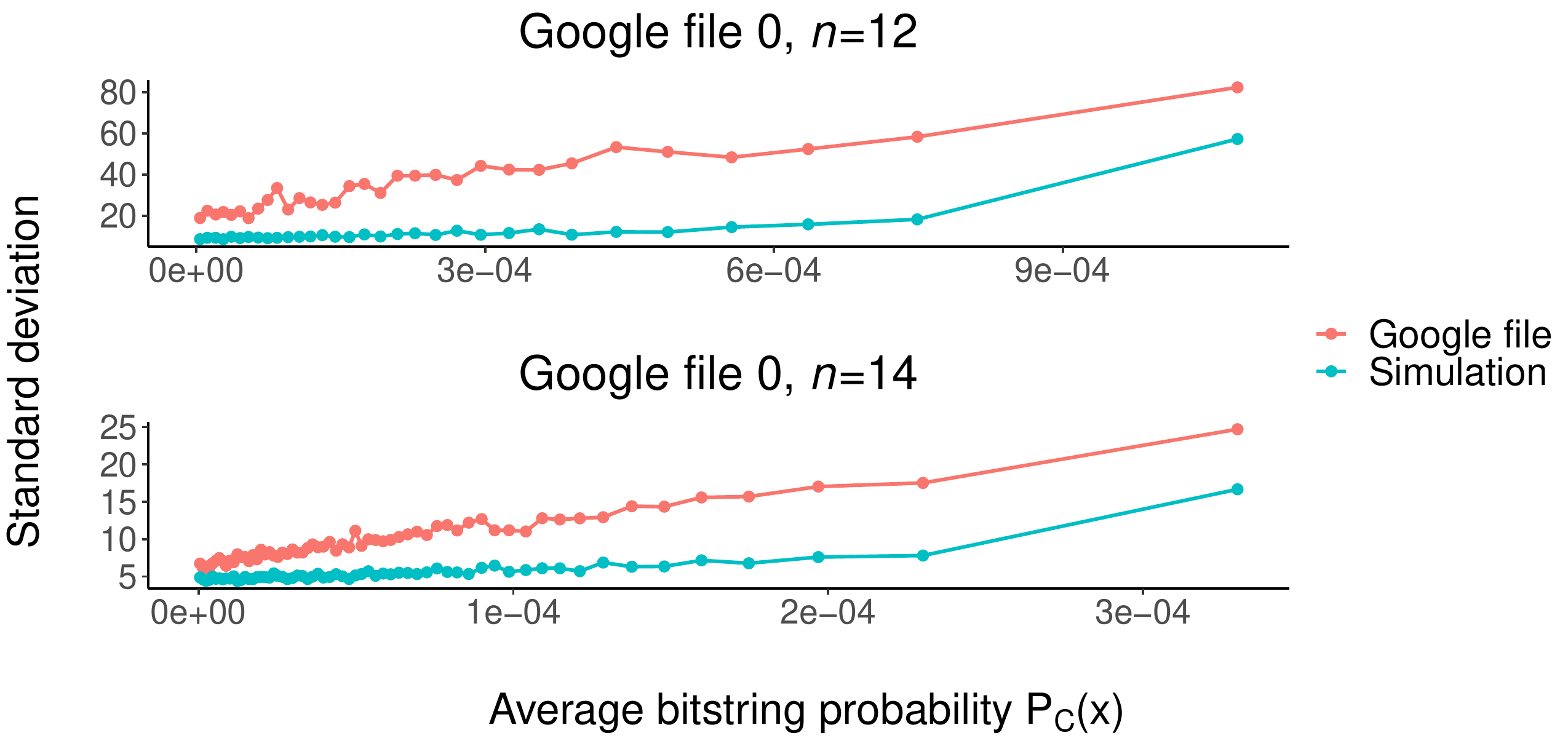}
\caption{We divided the bitstrings into groups of size 128 ($n=12$), and 256 ($n=14$) according to the values of ${\bf P}_C(x)$ and in each group computed the standard deviation of the empirical distribution in this group. The blue line represent the behavior for simulations based on Google's noise model with $\phi=0.3701$ for $n=12$, and $\phi=0.3298$ for $n=14$. The red line represents the behavior of the Google data.}
\label{int-var}
\end{figure}

\begin{figure}
\centering
\includegraphics[width=\textwidth]{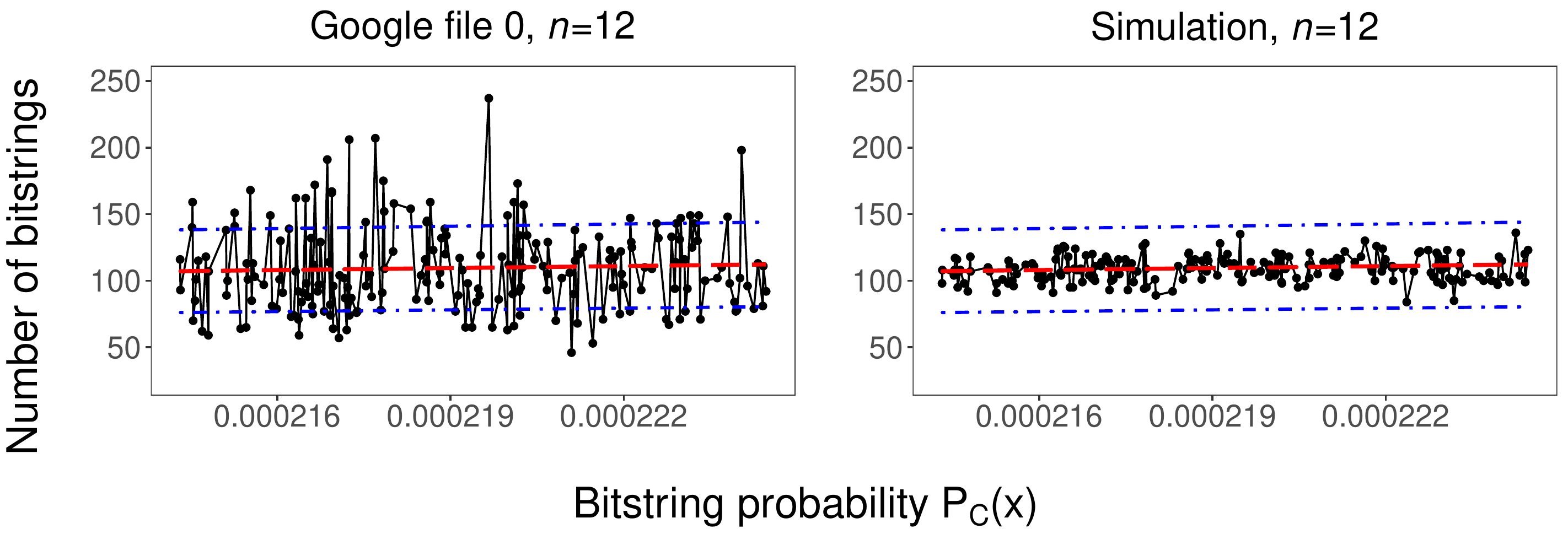}
\caption{  
Zooming in on the empirical frequency of bitstrings with amplitudes between the median and the 0.55 quantile. The left plot  is the empirical occurrences of the bitstrings in Google file 0, n=12. The right plot is based on a simulation with $\phi=0.3862$. The red line describes the expected number of bitstrings of the Google noise model,  
and the blue dashed line is 3 standard deviations from the expectation.} 
\label{fine-graned}
\end{figure}

We ordered all bitstrings by the value of $P_C(x_i)$, and divided the bitstrings into $s$ groups according to the value of $P_C(x_i)$ and considered the number of bitstrings corresponding to each group. 
For example, if $s=100$ we consider bitstrings in the sample with 
$P_C(x_i)$ in the top 1\% or between 10\% and 11\% from the top,
or between 49\% and 50\% from the top, and then these numbers agree perfectly with the Google model. Namely, they  equal, 
up to sampling errors, to the behavior for samples drawn by simulation from the Google model. However, when we considered the empirical variance for these bitstrings, it was very large (far beyond the effect of sampling). This is demonstrated in Figure \ref {int-var}.
In Figure \ref {fine-graned} we zoom in, for 12 qubits and Google's first bitstring file (file 0), 
on bitstrings $x$ for which the value 
of ${\bf P}_C(x)$ is between the median and the 0.55 quantile. (There are around 200 such bitstrings.) We compared the frequencies of appearances of every bitstring in the Google bitstring file with a file obtained by simulation.

\subsection {A detailed study of the empirical distribution}
\begin{figure}
\centering
\includegraphics[width=\textwidth]{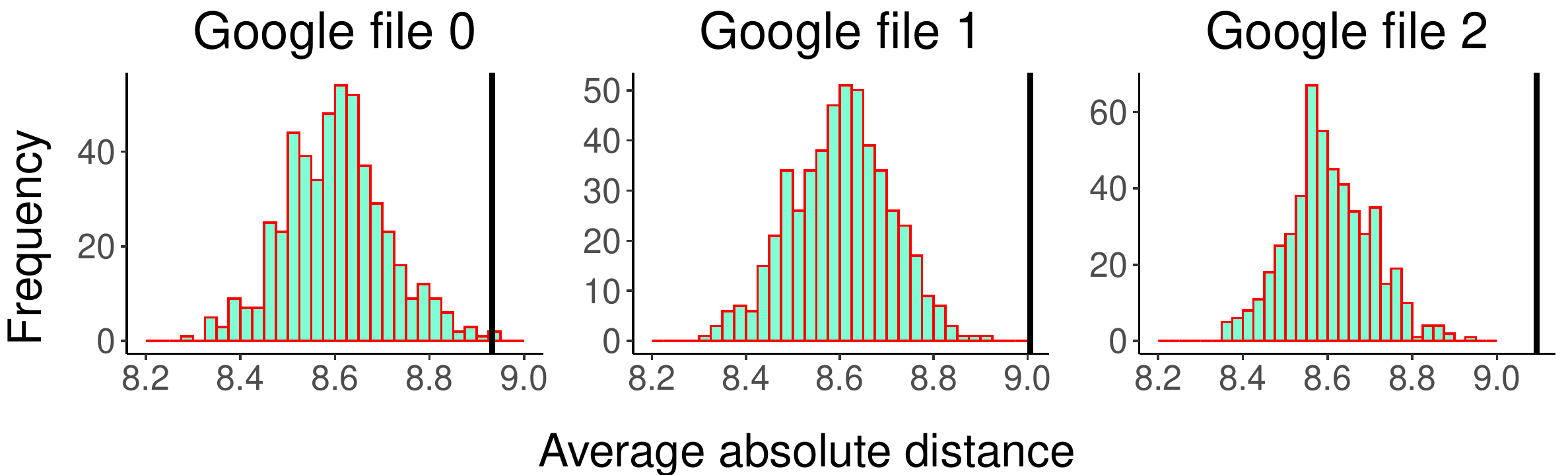}
\caption{Comparing the two halves of the Google samples: the 
black vertical lines  
are the $\ell_1$ distance of the occurrences of bitstrings when we partition the samples into two halves according to 
the sampling order. The histograms give the $\ell_1$ distances between the occurrences of bitstrings for random partitions of the bitstrings into two halves.}
\label{NS}
\end{figure}

\subsubsection {The non-stationary nature of the distribution}
\label {s:chaos}

\begin{figure}
\centering
\includegraphics[width=\textwidth]{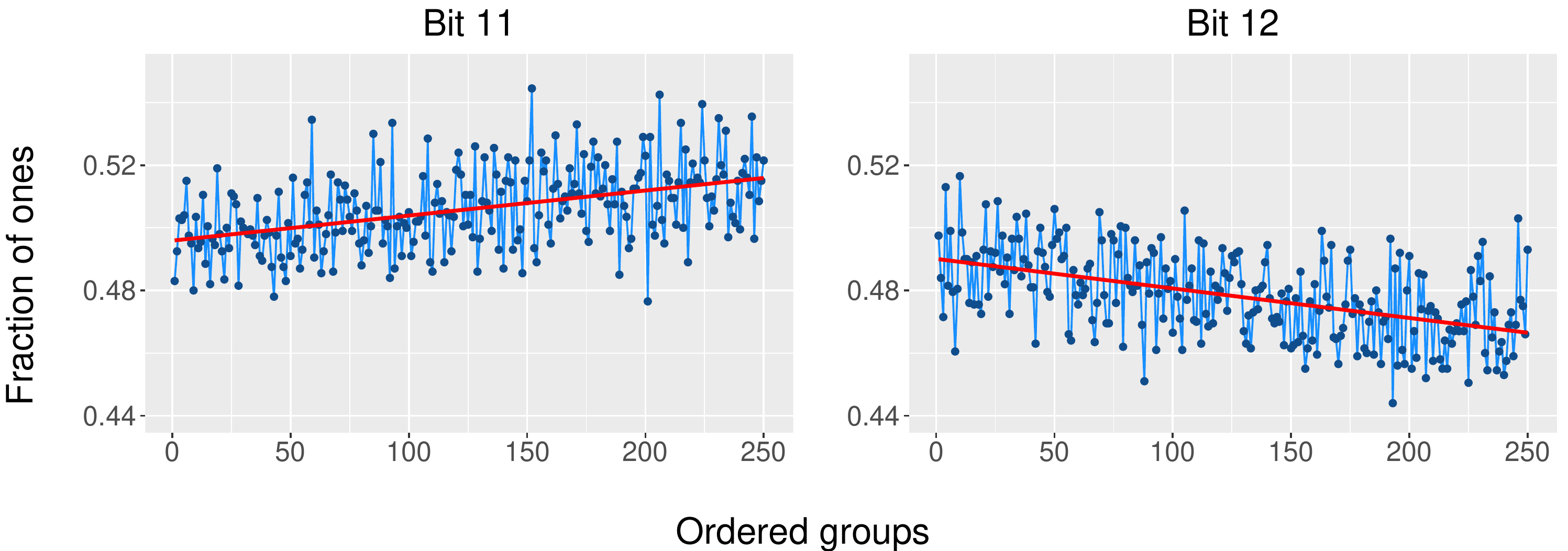}
\caption {Drifts in the fractions of ones. We divided the 500,000 bitstrings into 250 groups of 2,000 bitstrings each, according to the sampling order. For each group calculated the fraction of bitstring having a ``1" bit in some place in the bitstring. The Figure shows the fractions of ones in locations 11 and 12 for one circuit (file 0) with $n=12$. (The red lines are linear regression fits to the
data points.) The trend is consistent along the different circuits and is different for different locations in the bitstrings.}
\label{ones}
\end{figure}

For $n=12$ and for each bitstring file representing a sample of size 500,000 of the quantum computer, we divided the sample at random into two halves and computed the $\ell_1$ distance 
between the occurrences of bitstrings in the two halves. We compared these $\ell_1$ distances with the $\ell_1$ distance between the first and second halves of the bitstrings in their sampling order.
In Figure \ref {NS} we find that the distance between the two empirical distribution according to the sampling order is 
significantly larger than for random partitions. This finding indicates a non-stationary behavior of the samples. This behavior may be related to a stronger property, namely that a large component in the empirical distribution is noise sensitive. (Noise sensitivity and subsequent unpredictability of the data are predicted in Kalai's works \cite {Kal20,Kal18}, and go back to Kalai and Kindler \cite {KalKin14}.)
We observed some specific non-stationary behavior regarding the percentage of ``1'' for certain bits 
as a function of the sampling order (Figure \ref{ones}).

One would expect that these findings should translate to a reduction in the fidelity, but we did not find significant difference in the ${\cal F}_{XEB}$-fidelity (or other measures of fidelity) between the two halves of the data. We note also that recent experiments with the Sycamore quantum computer \cite {McE+21,Ach+23} exhibited periods in the data where the error rate jumped to much higher values and then decayed back to normal. (This may have been caused by cosmic rays.) In contrast we did not witness periods in the data with significantly lower ${\cal F}_{XEB}$-fidelity (for $n=12$).

\subsubsection {The apparent asymmetry}
\label{s:asym}
\begin{figure}
\centering
\includegraphics[width=\textwidth]{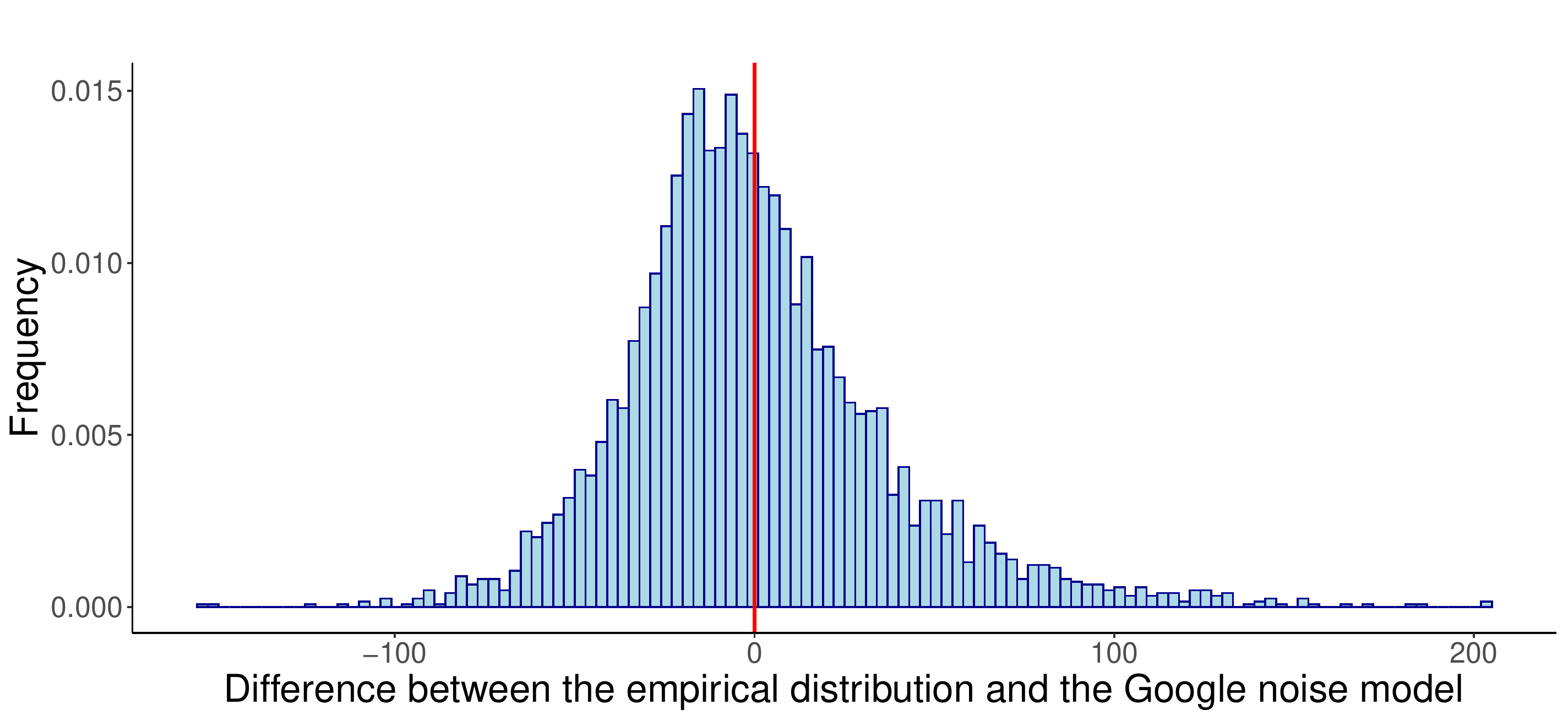}
\caption{A histogram of differences between the empirical distribution and the values given by the Google noise model, $n=12$, Google file 0, $\phi=0.3701$. (The behavior for other Google files was similar.)}
\label{asym2}
\end{figure}

When we compare the empirical distribution to the Google noise model 
we observe some asymmetry:  
the positive values of $O_C(x)-{\bf N}_C(x)\cdot N$ tend to be larger in absolute value than the negative values. This can be observed by looking at the points above and below the diagonal on the left side of Figure \ref {fig:scattt}. 
Figure \ref {asym2} shows that the  
distribution of the values $O_C(x)-{\bf N}_C(x)\cdot N$ is asymmetric. What can explain the apparent asymmetry in the gaps between the empirical distribution and the model? Is an explanation at all necessary?

With our parameters, the deviation due to sample errors of a sample based on the Google noise model from the distribution is symmetric (the right side of Figure \ref {fig:scattt}).  
We can expect that more accurate noise models will lead to a mixture of the correct Porter--Thomas distribution and a large number of uncorrelated Porter--Thomas distributions. This picture suggests that the individual differences $O_C(x)- {\bf N}_C(x)\cdot N$ behave like a Gaussian distribution and hence is symmetric. This can be further tested by simulations. 

\subsection {The Google team's response to the discrepancy between the data and the model}

The Google team regarded all along their noise model as a very simplified model of a very complex experiment, and asserted that ``it works well often, but not always." They emphasized the fact (referring to Section IV of \cite{Aru+19S}) that ${\cal F}_{XEB}$ does not rely on Google's specific noise model and works in much greater generality. (Indeed this is also confirmed by our paper \cite {RSK22}.) They  
raised the possibility that the Pearson chi-square test for circuits with 12 qubits in Section 8 of \cite {RSK22} fails because of the readout bias, and added: ``The bias can be seen by eye if you group probabilities by Hamming weight.
But this might not be enough. One might also try a simulation with noise, which is not that hard for 12 qubits. In any case, matching individual experimental probabilities with a simple noise model is not easy.'' 
They proposed to progressively study more realistic noise simulations in order to understand the apparent asymmetry. 
(In 2021, Google introduced a simulator for noisy circuits that may allow the implementation of this proposal, and noisy simulators were also developed by IBM, NASA, and others.) 

Regarding the excessive empirical variance, the Google team noted that they had a similar statistical study for a similar parameter
(referred to as {\it speckle purity benchmarking} (SPB), 
see Section VI.C.4 of  \cite {Aru+19S}), for small 1- and 2- qubit circuits, but they did not extend this study to large circuits.  The Google team 
also noted that our finding that the empirical behavior of the samples is not stationary is 
related to their studies in Sections VIII, H, and I of \cite {Aru+19S}.


\subsection {Modeling the noise: Readout errors and Fourier analysis}
\label {s:readout}
A general model for the samples can be described as follows: 
\begin {equation}
\label {e:nm}
{\bf N}^{*}_C =  \phi {\bf P}_C + \phi_{ro}{\bf N}_{ro} + (1-\phi_g){\bf N}_{g} + {\bf N}_T.
\end {equation}

The term $\phi {\bf P}_C$ is the ideal distribution described by the circuit $C$ multiplied by the fidelity, and this is the ``primary signal" we want to detect in the data.
The term ${\bf N}_{ro}$ represents the effect of readout errors conditioned on no gate errors. $\phi_g$ denotes the probability of not having any gate errors and $\phi_{ro}=:\phi_g-\phi$.  ${\bf N}_g$ denotes the effect of both gate and readout errors when there are gate errors. 
The term   ${\bf N}_T(x)$ corresponds to an additional substantial but poorly understood form of noise. (It can be regarded as some large fluctuation (with zero expectation) of the first three terms.) 
It is reasonable to assume that ${\bf N}_T$ is uncorrelated with ${\bf P}_C$ and 
 the other terms. The statistical estimators for the parameter $\phi$ under
the Google noise model \eqref {e:gnm} continues to apply under a more detailed model of the form \eqref{e:nm} (and this is already emphasized in \cite {Aru+19S}). 
In \cite {RSK22} (Section 6) we proposed two models for the readout errors and corresponding estimators for the overall effect of readout errors based on the assumption that the effect of gate errors is statistically independent of the effect of readout errors. In a subsequent work in progress \cite {KRS22f} we develop Fourier tools that allow quick computation of these estimators, and we also refine the fidelity estimators according to the Fourier levels.\footnote {For earlier studies of Fourier--Walsh expansion in the context of random circuit sampling see Boixo et al. \cite {BSN17}, Gao and Duan \cite {GaoDua18}, and Kalai \cite {Kal18}. See also Kalai and Kindler \cite {KalKin14} for an earlier related work on boson sampling. Aharonov et al. \cite {AGLLV22} is a recent relevant paper.}

\subsection {Further issues}

\subsubsection {Various classes of noise models}
\label {s:vnm}
To understand certain phenomena in the Google data, we also considered several large classes of noise models. Here is one example (from \cite {RSK22}): let ${\bf Q}$ be
a probability distribution on bitstrings of length $n$, and consider $2^n$ positive real numbers $\gamma_x$ for every bitstring $x$. Next consider ${\bf \tilde Q}(x)=\gamma_x{\bf Q}(x)$ normalized so that ${\bf \tilde Q}$ is a probability distribution. A special case is when we start 
with a probability distribution $D$ over $[0,1]$ and we choose at random  $\gamma_x$ according to $D$ for every bitstring $x$, where these choices are statistically independent.
When we let ${\bf Q}(x)={\bf P}_C(x)$ and apply this procedure we obtain a very different distribution with the same size-biased distribution, which demonstrates the asymmetry (Section \ref {s:asym}) 
and the behavior described in Figure \ref {int-var}. (We do not have a justification for this type of model for NISQ systems.)

\subsubsection {The data from patch circuits}

Patch circuits are simplified circuits that are composed of two distinct non-interacting parts. (The bitstrings and the circuit description for patch circuits were made available by the Google team in June 2022.) The data of patch circuits provided a new opportunity to compare the empirical distribution with the ideal probabilities and with various noise models already for 6-qubit circuits. A preliminary study of the data showed similar properties to those described in this section. 

\subsubsection {Other ways to estimate the fidelity}
The fact that the experimental data does not fit the Google noise model once again raises the interesting issue of estimating the fidelity. One way to estimate the fidelity is to consider the distance $D$ between the empirical distribution and the ideal distribution as the basic parameter, and to use the Google noise model as the basis
for the translation of $D$ into a fidelity estimator. (This is possible only for $n=12,14$.) We tried various notions of distance between distributions and (unsurprisingly) this gives lower fidelity estimations compared to those using linear cross-entropy. 

\subsubsection {Predicting the noise}
We ask the following question: given the values of ${\bf P}_C$, suppose that we want to predict the effect of the noise. For example, we want, based on the first half of the bitstrings, to predict the behavior of the second half, or based on bitstrings of one circuit, to predict the behavior if the bitstrings of another circuit. Of course such predictions are not possible for samples drawn according to the Google noise model since the difference between the two parts represents sampling errors. We
were not able to detect ways to predict the behavior of a circuit based on another circuit, but we were to (partially) predict the second half of a bitstring based on the first half.



\subsection {Summary of Section \ref{s:dist}}

There is a large gap between the samples of the Google quantum supremacy experiments and the Google noise model. In fact, the samples are far away from any noise model we are aware of. There is  evidence 
that the distance between the Google noise model and uniform distribution is smaller (when the number of qubits is $n>16$) than the distance between the experimental samples and the Google noise model. We studied other properties of the empirical distribution like its behavior in different scales, its non-stationary nature, and its Fourier behavior, and there is more to be done mainly for data coming from other NISQ experiments and data from simulators of noisy circuits.  

\newpage

\section {Concern II: Formula (77)}
\label {s:77}

\subsection {The predictive power of Formula (77)}
\label {s:77a}
As mentioned before, Formula (77) in the Google paper \cite {Aru+19S} (Equation (\ref{e:77}) in Section \ref {s:g1})
provides an estimation for the fidelity of a circuit based on the fidelities of its components:

$$\hat \phi~=~ \prod_{g \in {\cal G}_1} (1-e_g) \prod_{g \in {\cal G}_2} (1-e_g) \prod_{e \in {\cal Q}} (1-e_q).$$

The Google paper claims that this formula estimates with a precision of 10\%--20\% the
fidelity $\phi$ of the circuit identified as the probability of no errors ($p_{{\bf no~err}}$)  
of a circuit. This remarkable agreement is considered in \cite {Aru+19} and \cite {Aru+19S}
as an indication that there is no additional decoherence physics when the system scales.
It is considered a major new scientific discovery on its own. 
Our concern is that the predictive power of 
this formula, which implies no interaction between errors, and suggests complete independence, and no common (to many qubits)  environmental effect is surprising.
We discuss this concern in this section. It was raised in \cite {Kal22} (and also in \cite {Kal20v,Ira19} and earlier discussions in scientific blogs), 
and is mentioned in \cite {RSK22,KRS22d}.   

{\bf Remarks:} 
1) So far, the individual values of $e_q$ and $e_g$ in Formula (77) have not been made public.

2) An approximation to Formula (\ref {e:77}) based on averaged fidelities is given by 

\begin {equation}
\label {e:77s}
{\hat \phi}^{*}= (1-0.0016)^{|{\cal G}_1|} (1-0.0062)^{|{\cal G}_2|} (1-0.038)^n .
\end {equation}

3) The Google team proposed the following approximation that does not take into account the 1-gates, but adds their contributions to a combined fidelity for the 2-gates (referred to as 
2-gate cycles):
\begin {equation}
\label {e:77s2}
{\hat \phi}^{**}=  (1-0.0093)^{|{\cal G}_2|} (1-0.038)^n .
\end {equation}

Formula \eqref{e:77s2} can be seen as applying Formula \eqref{e:77s} under the assumption that the number of 1-gates is roughly twice the number of 2-gates, whereas in reality the number of 1-gates is considerably larger. 
Table \ref {T:77} compares the prediction from Formula (77) (as reported by Google), the simplified predictions described above, and the average fidelity for the experimental circuits. As expected, Formula (77) gives a better prediction than Formulas \eqref{e:77s} and \eqref{e:77s2}. 
Formula \eqref{e:77s2} 
gives considerably better predictions compared to Formula \eqref{e:77s}. 



\begin{figure}[h]
	\begin{center}
		\includegraphics[width=1.0\textwidth]{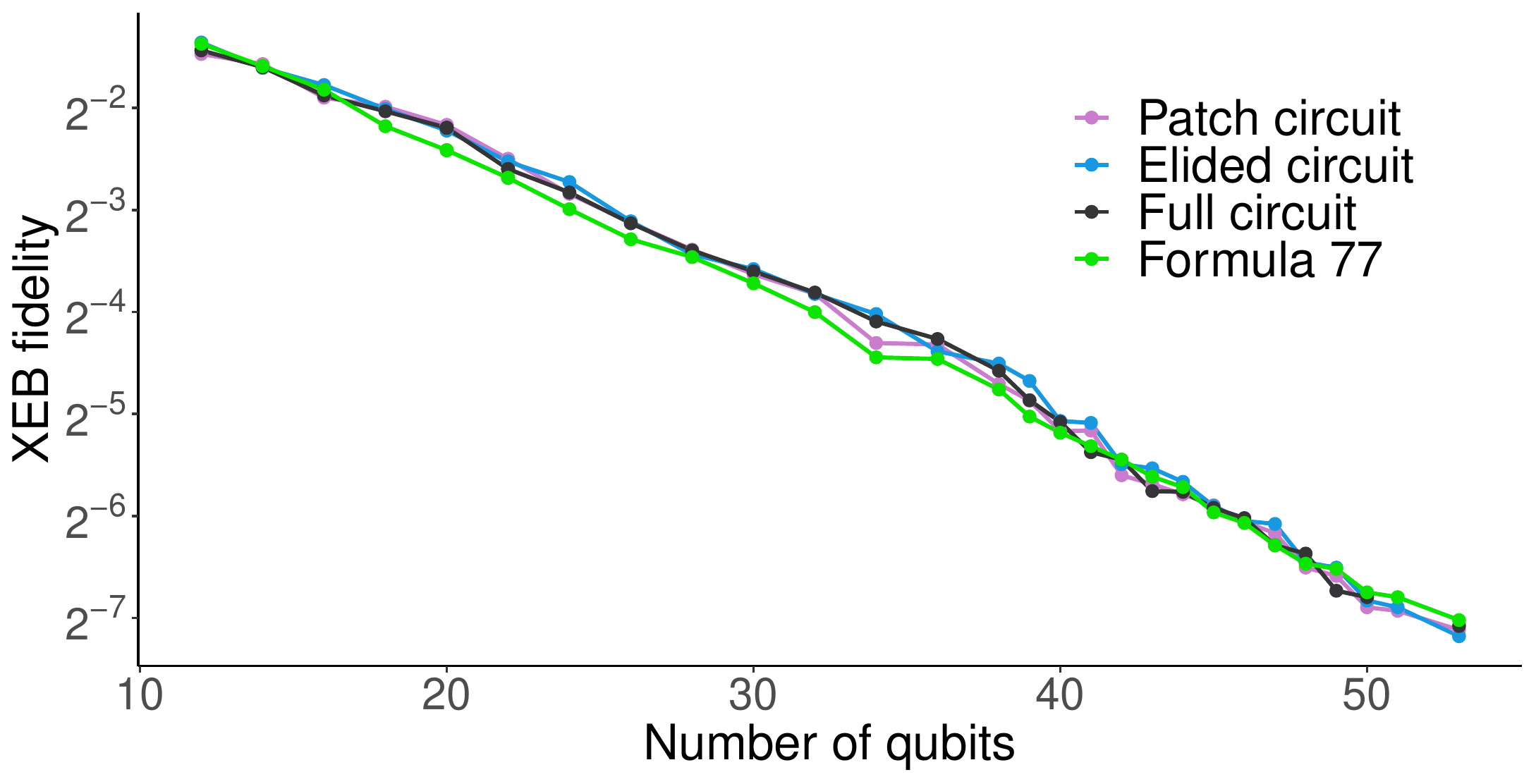}
		\caption{Comparison between average XEB fidelity for full, elided, and patch circuits with (77). Data from Table \ref{T:fidelity_all_circuits}}
		\label{fig:S28}
	\end{center}
\end{figure}

\subsection {The Google team's interpretation and statistical justification}
\subsubsection {Interpretation}

The Google paper \cite {Aru+19} regarded the predictive power of Formula (77) 
as a demonstration that there is no additional decoherence physics 
when the system scales. Here is a quote from the Google paper \cite {Aru+19S}

\begin {quotation} ``It is evident that there is a good agreement
between the measured and predicted fidelities, with deviations of up to only 10-20\%. 
Given that the sequence
here involves tens of qubits and $\sim$ 1000 quantum gates,
this level of agreement provides strong evidence to the
validity of the digital error model.
This conclusion can be further strengthened by the
close agreement between the fidelities of full circuits,
patch circuits, and elided circuits. Even though these
three methods differ only slightly in the gate sequence,
they can result in systems with drastically different levels
of computational complexity and entanglement between
subsystems. The agreement between the fidelities measured by these different methods, as well as the agreement
with the predicted fidelity from individual gates, gives
compelling evidence confirming the assumptions made
by the digital error model. Moreover, these assumptions
remain valid even in the presence of quantum entanglement.
The validation of the digital error model has crucial
consequences, in particular for quantum error correction.
The absence of space or time correlations in quantum
noise has been a commonly assumed property in quantum error correction since the very first paper on the
topic. 
Our data is evidence that such a property is
achievable with existing quantum processors."
\end {quotation}


Moreover, the Google team 
regarded the predictive power of Formula (77) as an important and amazing 
discovery. 
Here is a quote by John Martinis \cite {Mar19v} from November 1,  2019: 

\begin {quotation} 

``Now I want to point out what is the most amazing thing about this
data and why, if you want to think about it in a funny way, why we, quantum scientists,
can all keep our jobs or maybe get more funding (even better). And this is the data, look at this black line here...
It is just a simple prediction from the one- and two-qubit simultaneous data that I showed
you before... You multiplied all these fidelities together in a totally classical way, this is 
high school probability theory, and it predicts what is going on
in this very very complicated system with all this quantum computation going on. That, to me,
is absolutely amazing that
such a simple model for errors works so well.  When we got the data, when that happened, it
was the big surprise here. And, of course, if things were worse I wouldn't be standing here talking to you today.'' 
\end {quotation}

In the wake of efficient classical algorithms for computing amplitudes and the decline of  Google's quantum supremacy claim,
members of the Google team as well as other researchers 
regarded the
independence of errors that allows the predictive power of Formula (77) as the most important remaining
scientific achievement of the Google experiment, and viewed the statistical independence it exhibits as a strong support that the statistical independence assumptions for errors required by quantum fault tolerance are achieved.         

\subsubsection {Statistical justification} 

The Google team (November 2019) justified the remarkable predictive power of
their Formula (77) (Equation (\ref {e:77})) 
with a statistical computation that is based on the following three ingredients.

\begin {enumerate}
\item
Estimations for individual readout and gate error probabilities are accurate. 
The Google team reported that the error rate for their estimations of individual qubit and gate error probabilities is 20\%.
\item
Mistakes in the estimations of individual error probabilities are unbiased; namely, there are no systematic mistakes in these estimations.
\item
  Gate errors and readout errors are statistically independent.
  \end {enumerate}

Based on these assumptions,
    Google's (rough) estimation 
 of the relative deviation of the prediction of Formula (77) was 
 \begin {equation}
   \label {e:fn}
 DEV = 0.2 \cdot ( \sqrt n\cdot 0.038 + \sqrt{|{\cal G}_1|}\cdot 0.0016 + \sqrt{|{\cal G}_2|}\cdot 0.0063 ).
     \end {equation}
 For example,  for $n=53$ and $m=14$, the number of 1-gates $|{\cal G}_1|=795$ and the number of 2-gates $|{\cal G}_2|=301$, and Formula \ref {e:fn}  yielding, roughly 8.6\%. 
 
 The explanation for this estimation is that once we accept that the errors in assessing the actual fidelities are unbiased (or non-systematic), their cumulative effect grows like the square root of the number of components (as in simple random walks). For example, for qubit errors, the probability of an error in a single qubit is, on average, $0.038(1\pm 0.2)$. The cumulative effect of the $(1\pm 0.2)$ terms for $n$ qubits is $\sqrt n$ multiplied by 20\% of the average error, yielding $0.2 \cdot \sqrt n \cdot 0.038$.  
   

\subsection {Our view on the statistical assumptions behind Formula (77)}

In our opinion, the most unreasonable assumption behind Formula (77) is the statistical independence assumption. It is unreasonable to assume statistical independence for the failure of components in a quantum computer with hundreds and even thousands of components. 
Next, let us consider the issue of biased versus unbiased errors.  The following computation shows that even a small bias can have a large effect on the fidelity.
Fix a circuit $C$. Suppose that 
for every component of $C$ the error rate is 20\% larger with probability $(0.5+t)$ and 20\% smaller with probability $(0.5-t)$.  The deviation of Formula (77) (when $t$ is small) 
will be given by: 
$$  (1-2t)DEV+ 2t\cdot (n \cdot 0.038 +|{\cal G}_1|\cdot0.0016  +|{\cal G}_2| \cdot 0.0063) ,$$ 
where $DEV$ is the value in Formula \ref {e:fn}.
For $n=53, m=14$, and $t=0.05$ this gives roughly 50\%. 
Finally we note that the non-stationary behavior that we studies in Section 
\ref {s:chaos} may weaken the statistical justification of Formula (77). Specifically, the drifts in the fraction of ones might be in tension with both the $\pm 20 \%$ estimate and the non-bias assumption.

{\bf Remarks:} 1) The methodology of \cite {Aru+19} already assumes some sort of dependence between errors. Namely, for every component of the computer we have one error estimation when the component works in isolation, and a higher error estimation when the component works as part of a larger computation; these higher error estimates are used for Formula (77). (It is reasonable to assume that higher error estimation would apply for components of  larger circuits, but this is not the case.) As we already mentioned we do not have the data regarding failure probabilities of the different components and,  in particular, not the statistical data behind the claim of $\pm 20\%$ errors in the error estimation themselves.   

2) We note that positive correlation between the errors   
will actually lead to higher fidelity.

3) The three points in the Google statistical justification for formula (77) assume that 
${\cal F}_{XEB}$ is a good estimator for $p_{no~err}$.  
In contrast, Gao et al. \cite {Gao+21} showed (see also Section \ref {s:gao}) that there is a systematic difference
and $${\cal F}_{XEB} > p_{no~err}.$$ 
There are various reasons for this gap and one of them is that two errors can cancel and lead to an effect of ``no errors."   Gao et al. expected in the paper that 
for the Google and USTC circuits 
``XEB values would overestimate fidelities only by a few percents," but this expected gap deserves further understanding. A large difference between ${\cal F}_{XEB}$ and $p_{no~err}$ 
would provide an additional reason to doubt Formula (77). It will thus be interesting to further study the difference between ${\cal F}_{XEB}$ and $p_{no~err}$ using simulations.

\subsection {Other responses to our concern}

Several researchers offered various possible explanations for the predictive power of Formula (77) (and the required statistical independence):
one explanation for the success of Formula (77), suggested by 
Peter Shor (in a blog discussion \cite {Sho19}) 
Boaz Barak in \cite{Ira19}, 
and others, is that 
the statistical independence needed for the success of Formula (77) 
is justified in a typical situation for random circuits. We do not see a justification 
for this claim and note that the prediction of Formula (77) is very good also for circuits that are not random (such as patch and elided circuits). 

Adam Bouland (private communication, 2020) offered an explicit model that shows that Formula (77) is also consistent with various forms of correlated noise: when a specific 1-gate is faulty, a random Pauli operator applies to all qubits in the circuits rather than only to the gated qubit. 
Bouland added: ``So I think a more reasonable criticism of their result is that their metric is too generous (as it hides correlations in noise), rather than that the formula is too good to be true (which I find difficult to claim without questioning the validity of their experimental results, which would be a serious accusation)." 
We agree that Formula (77) supports also various models of correlated noise but we do not think that  
this remark has a direct bearing on our concern regarding the predictive power of Formula (77).
While it is correct to view the statistical independence manifested by the predictive power of (77) as a source of optimism for quantum fault tolerance, Bouland's example shows that this is not sufficient to exclude certain correlations that may cause quantum
fault-tolerance to fail (nor it is necessary).

Greg Kuperberg (private communication, 
2020) made the distinction between systems 
with moving parts (like an airplane)
and systems without moving parts (like the Sycamore chip),
and asserted that our concern would have been justified for
systems with moving parts but not for the Sycamore chip, which is ``the size of a thumbnail, has no moving parts (save for electrons), and is shielded by a dilution refrigerator." 

Several researchers  have raised the point that the concern about Formula (77) amounts to 
challenging the integrity of the Google data and experiment.  
For example, Adam Bouland and Dorit Aharonov raised this point in \cite {Kal20v}.  
Aaronson \cite {Aar20} opined that the concern about the predictive power of Formula (77) amounts to suggesting that the Google data had been tampered with or faked, but added that the successful predictions based on Formula (77) are exactly what a good experiment should have produced!  

We note that in this paper we take the position that concern about Formula (77) or feeling surprised by it does not amount to an accusation.

\subsection {Confirmations and  replications}

\subsubsection {Confirmation of the predictive power of Formula (77) by Kalachev,
  Panteleev, and Yung}

Considerable 
support for Google's fidelity claim, and the predictive power of Google's Formula (77), 
came from a recent work
of Kalachev, Panteleev, and Yung 
\cite {KPY21}. They presented improved algorithms for amplitude computations 
and computed the amplitudes for samples of almost two hundred circuits ($38 \le n \le 53$) for which the Google team did not compute the amplitudes. In all these cases, the ${\cal F}_{XEB}$ perfectly agrees with Formula (77) prediction. 
A recent further remarkable confirmation was achieved by Liu et al. \cite {Liu+22} who managed to compute the amplitudes for one circuit out of Google's largest 53 qubits depth 20 circuits. The ${\cal F}_{XEB}$ fidelity agrees with Formula (77). (Of course, as mentioned in Section \ref {s:g1} these advances largely refuted Google's quantum supremacy claim.)  

These confirmations attenuate 
the concern about Formula (77) 
since  
its predictive power extended to those two hundred circuits for which the Google team stated that they had not even computed the amplitudes, and to dozens of circuits for which they were not able to compute the amplitudes at all.

\subsubsection {Replications and similar experiments}

USTC's close replications of the Google experiment \cite {Wu+21,Zhu+22} also provide independent 
support for the predictive power of Google's Formula (77). We did not study 
the data from these experiments and this remains an interesting challenge. 
With the exception of USTC's replications, the closest random circuit sampling experiment we are aware of is by  Kim et al. \cite {Kim+21}, a team of researchers 
from IBM.  The paper  
describes circuits with at most six qubits (and large depth). 
We could not find reports and data on random circuit sampling by quantum computers of other academic groups or companies.

\subsection {Formula (77) for the individual patches and the combined circuits}
\label {s:patch2}
Recall that patch circuits are obtained from full circuits by deleting the 2-gates that operate on qubits from both patches. Thus, each of the two patches can be regarded as a separate random circuit. The Google team provided the data for the patch circuits (except for those of 
type {\bf EFGH}, $n=53$, $m=14$, which are still unavailable) in June 2022. 
We computed the fidelities of the two patches for the patch circuits and observed
large gaps between the fidelities of patches. For example,
for $n=40$, the first patch has fidelity 0.139 while the second patch has fidelity 0.220. 
For the $n=53$, $m=20$ experiment, 
one part has fidelity 0.026 while the other part has fidelity 0.070. 
(See Tables \ref {T:77-patched_combined}, \ref {T:77-patched_ABCDCDAB_combined} 
and Figure \ref {fig:patch++} in the Appendix.)
By contrast, the approximations we use
based on averaged fidelities (Formulas (\ref {e:77s}) and (\ref{e:77s2})), lead to similar values for the two patches and therefore 
do not apply. 
For further study of this point we 
need the individual values for the qubit- and gate- fidelities that were used for computing the predictions based on Formula (77) that are reported in the paper. 

\subsection {A systematic deviation of Formula (77) for the patch circuits}
\label {s:patch1}

The agreement in linear cross-entropy estimator between patch, elided, and full circuits 
was required to establish the extrapolation argument in the Google paper, and the significance of this agreement was emphasized in \cite {Aru+19,Aru+19S} (e.g., Figure 4 in \cite {Aru+19} and the quote below).
However, another concern about Formula (77) (that was raised in the Google paper itself) is that 
while Formula (77) predicts that the ${\cal F}_{XEB}$ fidelity of patch circuits will be 10\% higher than the fidelity of full circuits, in reality, over hundreds of circuits, the fidelities of patch and full circuits will agree. 
The reason that Formula (77) predicts higher fidelity for patch circuits compared to full circuits is that patch circuits have a substantially smaller number of 2-gates, and the omitted 2-gates do not contribute to the product formula.

The following quote from Section VIII.B of \cite {Aru+19S} raises this issue:   

\begin {quotation}
``For every system size investigated, we found that patch and full XEB provide fidelities that are in good
agreement with each other, with a typical deviation of
$\sim$ 5\% of the fidelity itself (we attribute the worst-case
disagreement of 
10\% at 34 qubits due to a temporary
system fluctuation in between the two datasets, which
was also seen in interleaved measurement fidelity data).

Theoretically, one would expect patch circuits to result in
$\sim$ 10\% higher fidelity than full circuits due to the slightly
reduced gate count. We find that patch circuits perform
slightly worse than expected, which we believe is due to
the fact that the two-qubit gate unitaries are optimized
for full operation and not patch operation. In any case,
agreement between patch and full circuits shows that
patch circuits can be a good estimator for full circuits,
which is quite remarkable given the drastic difference in
entanglement generated by the two methods. These results give us a good preview of the system performance in all three regimes discussed earlier."
\end {quotation}

The authors did note that Formula (77) predicts that the fidelity for patch circuits will be higher (by roughly 10\%) and offered the explanation that the systematic violation of Formula (77) reflects ``optimization for full operation rather than patch operation." This explanation apparently refers to the calibration process and we do not find it convincing. We will elaborate on this matter in Section \ref {s:patch-cal}.

\subsection {Summary of Section \ref{s:77}}

The remarkable predictive power of Formula (77) is statistically surprising: the subsumed independence between components of the quantum computer,
is  striking.
The close agreement of the experimental ${\cal F}_{XEB}$ fidelities between the patch circuits and full circuits 
shows an unexplained systematic deviation from the predictions of Formula (77). 
On the other hand, confirmations \cite {KPY21, Liu+22} and replications \cite {Wu+21,Zhu+22} lend support to the claims in the Google paper.
There are various matters that remain to be explored. For example, (i) using simulations of noisy circuits, the magnitude of the difference between the two sides of Formula (77) (Gao et al. \cite {Gao+21}) could be estimated, and (ii) the individual values in Formula (77) could be used to study the different ${\cal F}_{XEB}$ fidelities of the two patches in patch circuits.  

\newpage

\section {Concern III: The calibration process} 
\label {s:calib}

\subsection {The calibration process and its local nature}

By the calibration process we refer to a method of central importance in Google's quantum supremacy experiment, which, based on multiple runs of
1-qubit and 2-qubit quantum circuits, adjusted the definition of the experimental circuit 
to the way the Sycamore quantum computer actually behaved. The calibration procedure was described in the Google paper (Section VI.B.4 of \cite {Aru+19S}) as  an optimization algorithm where 
the ``optimization search space [is] of [dimension] 
$\sim 100^{4N}$ [which is] much larger than the dimension of the
Hilbert space of an $N$-qubit processor, which is $2^N$."

To make matters clear,  the calibration is not about tightening the screws in Sycamore; rather, it is about change in the program.
We can think about the calibration process as a change in the model that would greatly reduce certain systematic
forms of noise. For example, if we discovered that a certain 1-gate that is supposed to apply a 90-degree rotation systematically performs an 80-degree rotation, rather than changing the engineering of the 1-gate, we would change the definition of the circuit.

The calibration consisted of certain modifications to every 2-gate $g$ of the circuits. If the 2-gate $g$ acts on qubits $x$ and $y$ then 
\begin {itemize}
\item
The parameters of the 2-gate $g$ were modified, and, in addition, 
\item
Two pairs of 1-gates  (referred in \cite {Aru+19S} as ``$Z$-rotations") were added to the definition of the circuit. Each pair had one gate operating on qubit $x$ and one gate operating on qubit $y$, with one pair being added before the 2-gate $g$ was executed and one pair being added afterwards. 
\end {itemize}

There were no additional modifications to 1-gates of the circuits. (See \cite {KRS22d} Section 3 for more details.)
These calibration modifications were carried out simultaneously in
all the experiments of pattern {\bf EFGH} on patch, elided, and full circuits. (A separate calibration process was 
run for the non-verifiable full circuit of pattern {\bf ABCDCDAB}.) 

The Google team explained to us that most of the $Z$-rotation (for a given 2-gate $g$) is to account for ``phase changes due to different idle frequencies of the interacting qubits," and it is determined from the fixed qubit frequencies and gate times. (Furthermore, this component has been removed in newer experiments.) 
Putting this component of the $Z$-rotation aside, the calibration method accounts for small systematic errors in the experimental circuits compared to the random circuits they represent. 

The calibration adjustments to a 2-gate $g$ that operates on qubits $x$ and $y$ were based 
on data from multiple runs of 1-qubit and 2-qubit circuits involving these two qubits. 
The multiple runs of 1-qubit and 2-qubit quantum circuits were not made in isolation but in parallel, and each run 
represented  layers of 2-gates of 
the same type, where, just like in the experimental circuits,  there were layers of random 1-gates 
between these 2-gate layers. 
We note that in applying layers of 2-gates of the same type along with random 1-gates between these layers, every qubit interacts with (at most) a single other qubit; therefore, we obtain many 2-qubit circuits (and 1-qubit circuits) running in parallel.  The calibration method was based on running these circuits with only one type of 2-gate layer, and making the adjustments to 2-gates so as to maximize the ${\cal F}_{XEB}$ value.  
It seems that the same algorithm 
can also be applied when the 2-gate layers are not the same and, in particular, to the experimental circuits themselves.

As mentioned in \cite {KRS22d}, the precise procedure that
the Google team used to move from the data gathered on small circuits (on 1- and 2- qubits) to the
list of modifications to every 2-gate is a commercial secret. 
We also do not have the experimental data for the small circuit experiments that consisted of the input to that process.
In \cite {KRS22d} we raised  several other non-statistical 
issues 
about the calibration process, and we mention two 
of them here. The first 
issue 
is that the calibration process weakens the claim for a ``programmable quantum computer.'' (This issue is supported by our statistical study here.) 
The second issue 
is that improvements of the calibration process were interlaced with the experiment, and, that the last minute calibration procedure for the {\bf EFGH} circuits (see Section 4.5 of \cite {KRS22d}) represented a substantial improvement.


The crucial property of the calibration procedure is its {\it locality} in the following strong sense: Rather than a general optimization process on a large number of parameters describing all the 2-gates of the circuit, the modifications for a 2-gate $g$ that operates on qubits $x$ and $y$ are based on runs of 1-qubit and 2-qubit circuits involving $x$ and $y$, and are primarily meant to cancel systematic noise for this particular 2-gate.  
In the rest of this section we present a statistical study of the calibration process, and share some statistical findings that are surprising, especially given the locality of the calibration.

\begin{figure}
\centering
\includegraphics[width=0.9\textwidth]{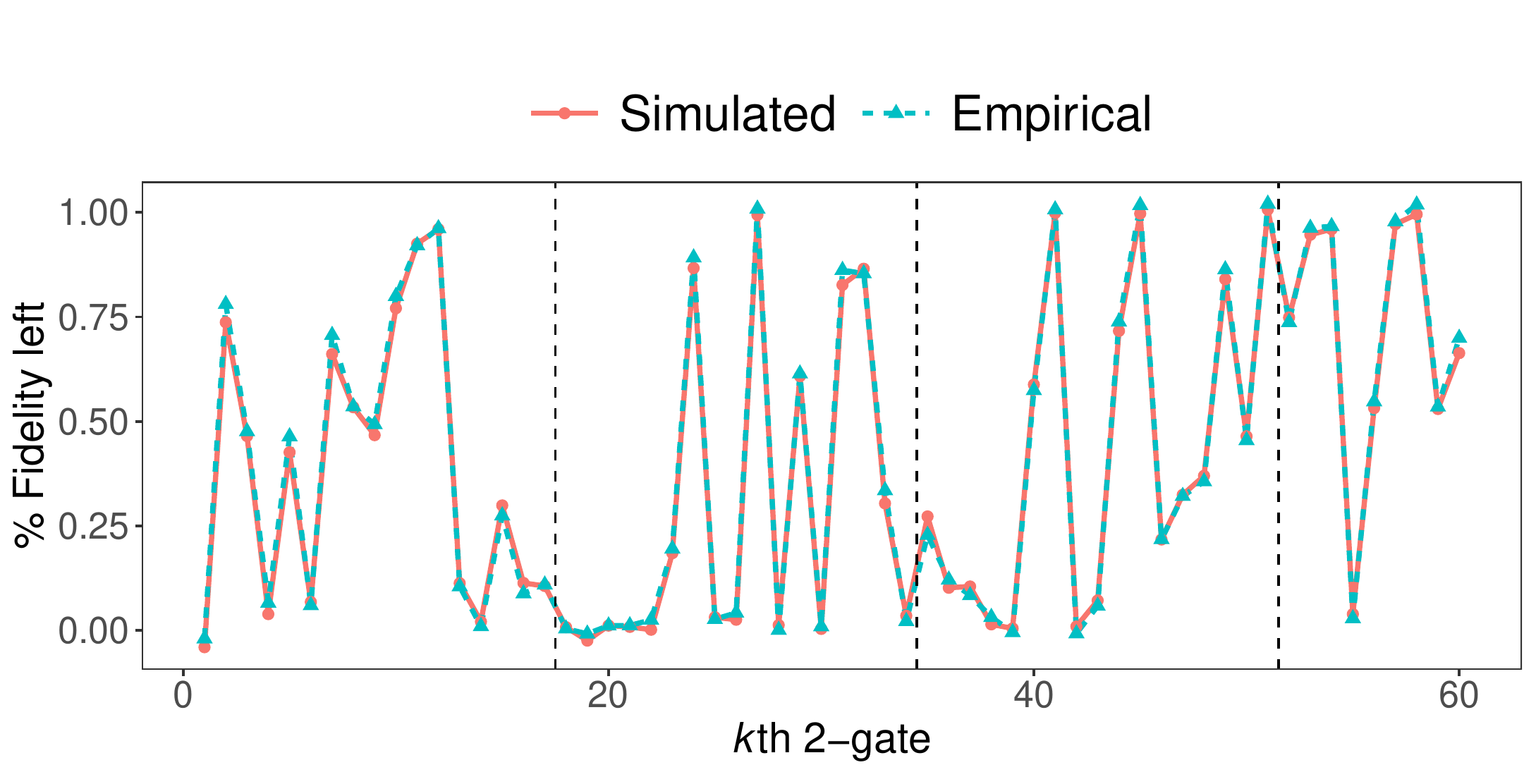}
\caption{The remarkable effectiveness of the calibration I. The effect of removing the calibration for the $k$th 2-gate of a circuit for Google's file 0 with $n=12$. Note that the same 2-gate occurs periodically along the circuit, indicated by the vertical black dashed lines. Here we remove all ingredients of the calibration: both the 1-gate rotations and 2-gate adjustments.} 
\label{calib-all}
\end{figure}

\subsection {The outcomes fully rely on the calibration process} 
\label {s:1gc}
When we start with an experimental random circuit
without making the 1-gate and 2-gate modification and compute the ${\cal F}_{XEB}$-fidelity of the experimental bitstrings,
we find that the fidelity 
is slashed to zero. 
Moreover, the 
effect on the fidelity is substantial even if we do not apply only the modifications
that correspond to one occurrence of a specific 2-gate: 
in many cases, the effect is substantial, in some cases it slashes the fidelity to being close to zero, and in some other cases it has small effect (Figure \ref {calib-all}). 
On this matter the Google team referred us to Section VII.E of \cite {Aru+19S}, where the systematic errors that are corrected by the four $Z$-rotations for the qubits of each 2-gate are explained:  ``The $Z$-rotations have two origins. First, they capture
the phase shifts due to qubit frequency excursions during the two-qubit gate. Second, they account for phase
changes due to different idle frequencies of the interacting qubits. The latter introduces dependency of the three
parameters defining the $Z$-rotations on the time at which
the gate is applied. By contrast, for a given qubit pair $\theta$
and $\phi$ do not depend on the cycle." 

\subsection {The calibration seems unreasonably effective}
\label {s:ue1}
In Figure \ref {calib-all} we also make a comparison between two quantities for every 2-gate among the 60 2-gates of the circuit. One quantity is the fraction of remaining fidelity when we do not apply the calibration for a single 2-gate. These fidelities are represented by the little blue triangles. The second quantity assumes that the calibration gives a perfect description of the circuit: we sample according 
to the the calibrated circuit  
and then compute the fidelity when the 2-gate calibration is cancelled. 
Those fidelities are described by the little red circles. 
The perfect agreements between the red line and blue dashed lines shows that the 2-gate calibrations are very effective, ``right on target." 
The calibration accounts for ``noise cancelling" for systematic errors on the individual 2-gates and it is surprising that the calibration process achieved a perfect noise canceling. 
Mathematically speaking, The property that improvements for each 2-gate is close to perfect means that the calibration process reaches (approximately)
a critical point of the function ${\cal F}_{XEB}$ over the many parameters of the different 2-gates. Reaching a critical point would be 
less surprising for a global optimization process for maximizing ${\cal F}_{XEB}$ over all these parameters simultaneously. 




\subsection {Google's explanation for the agreement in the ${\cal F}_{XEB}$ estimator
between patch and full circuits}
\label {s:patch-cal}
In Section \ref {s:patch1} we discussed
the concern that 
while Formula (77) predicts that the ${\cal F}_{XEB}$ fidelity of patch circuits will be 10\% higher than the fidelity of full circuits, in reality, over hundreds of circuits, the fidelities of patch and full circuits agree.  
The Google paper \cite {Aru+19S} offered the explanation that the systematic violation of Formula (77) reflects ``optimization for full operation rather than patch operation," apparently referring to the calibration process discussed in this section.  
The calibration for a 2-gate $G$ involving qubits $x$ and $y$  
is based on a multiple run of 1-qubit and 2-qubit circuits involving these two qubits. 
This is done 
by alternating between layers of 2-gates of the same type and (between them) layers of random 1-gates. The difference between ``full" operation and ``patch" operation is the presence of a handful of 2-gates that are not directly interacting with qubits $x$ and $y$, 
and it seems unreasonable that this difference will have any impact.

\subsection {2-gate adjustments seem unreasonably effective} 
\label {s:2gc}
 
\begin{figure}[h]
\centering
\includegraphics[width=\textwidth]{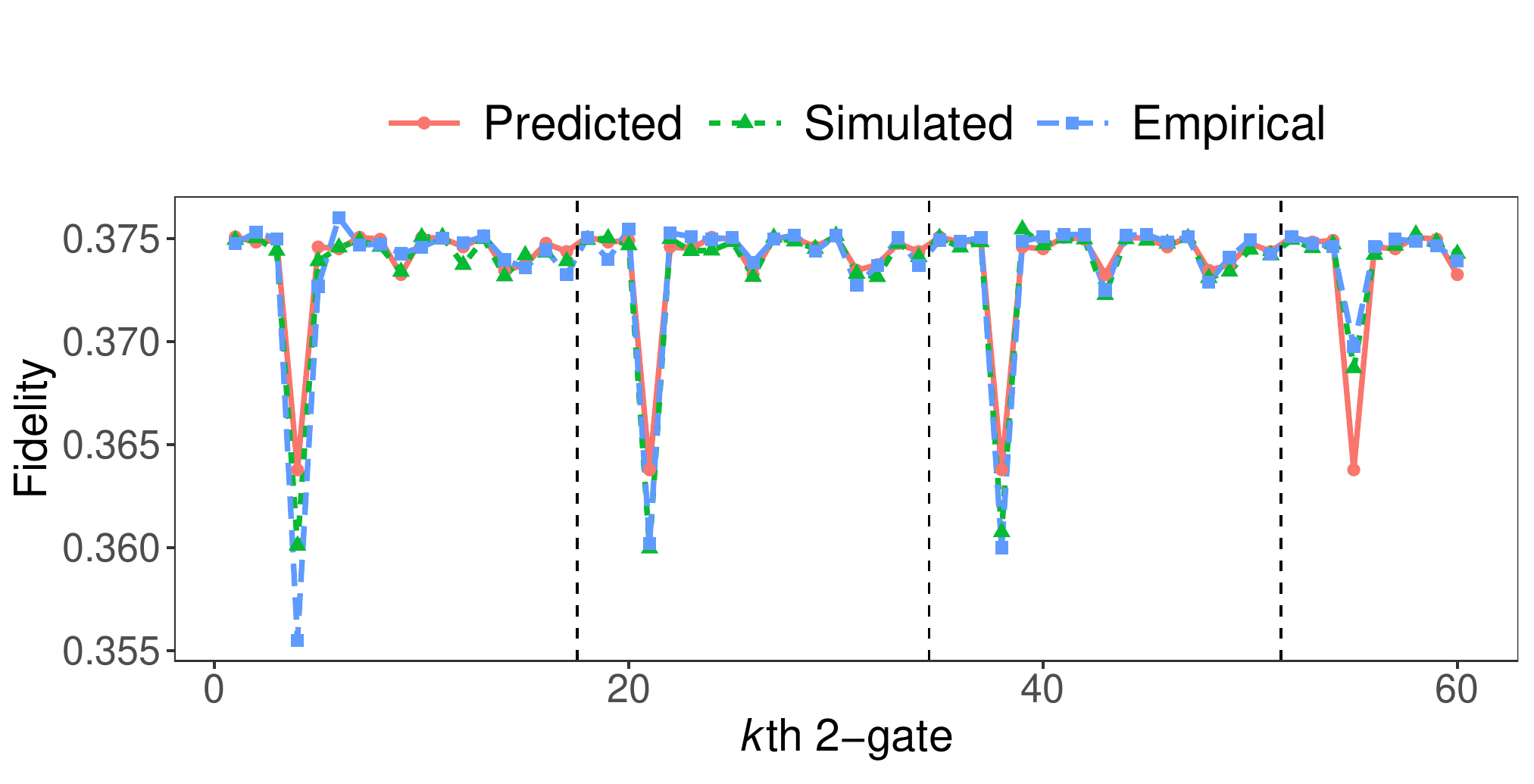} 
\caption{The remarkable effectiveness of the calibration II. The effect of removing the 2-gate adjustments involving the $k$th 2-gate of a circuit. (The same 2-gate occurs periodically along the circuit.) The 2-gate involving the qubits (3,3) and (3,4) has consistently large effect. 
\label {fig:2ga}
} 
\label{calib6}
\end{figure}

In this subsection we leave unchanged the $Z$-rotations 
that were added in the calibration process, and study the effect of the adjustments to the definition of the 2-gates.
The Google team asserted that for the experiment with $n=53$ and $n=14$, removing the 2-gate adjustments would slash the fidelity by a factor of two, and provided formulas \eqref{eq:red_fid_cal} and \eqref{eq:red_fid_tot} (below) 
to estimate this effect. 
In Figure \ref {fig:2ga}, the effect on the fidelity of removing the 2-gate adjustments of each of the 60 2-gates for the first experiment $n=12$ is represented by the blue squares. The small green triangles describe this same effect by simulations (50 simulations for every 2-gate), assuming that the adjustment is perfect. The red circles describe the effect by Formula \eqref{eq:red_fid_cal} below (that also assumes that the adjustment is perfect). The empirical effects are very close to the ones that assume that the adjustments are optimal and show that the calibration is ``right on target." In technical terms, when we consider ${\cal F}_{XEB}$ as a function of the parameters of the 2-gates, the parameters achieved by the calibration are close to be a critical point of the function.  

\begin{figure}
\centering
\includegraphics[scale=0.5]{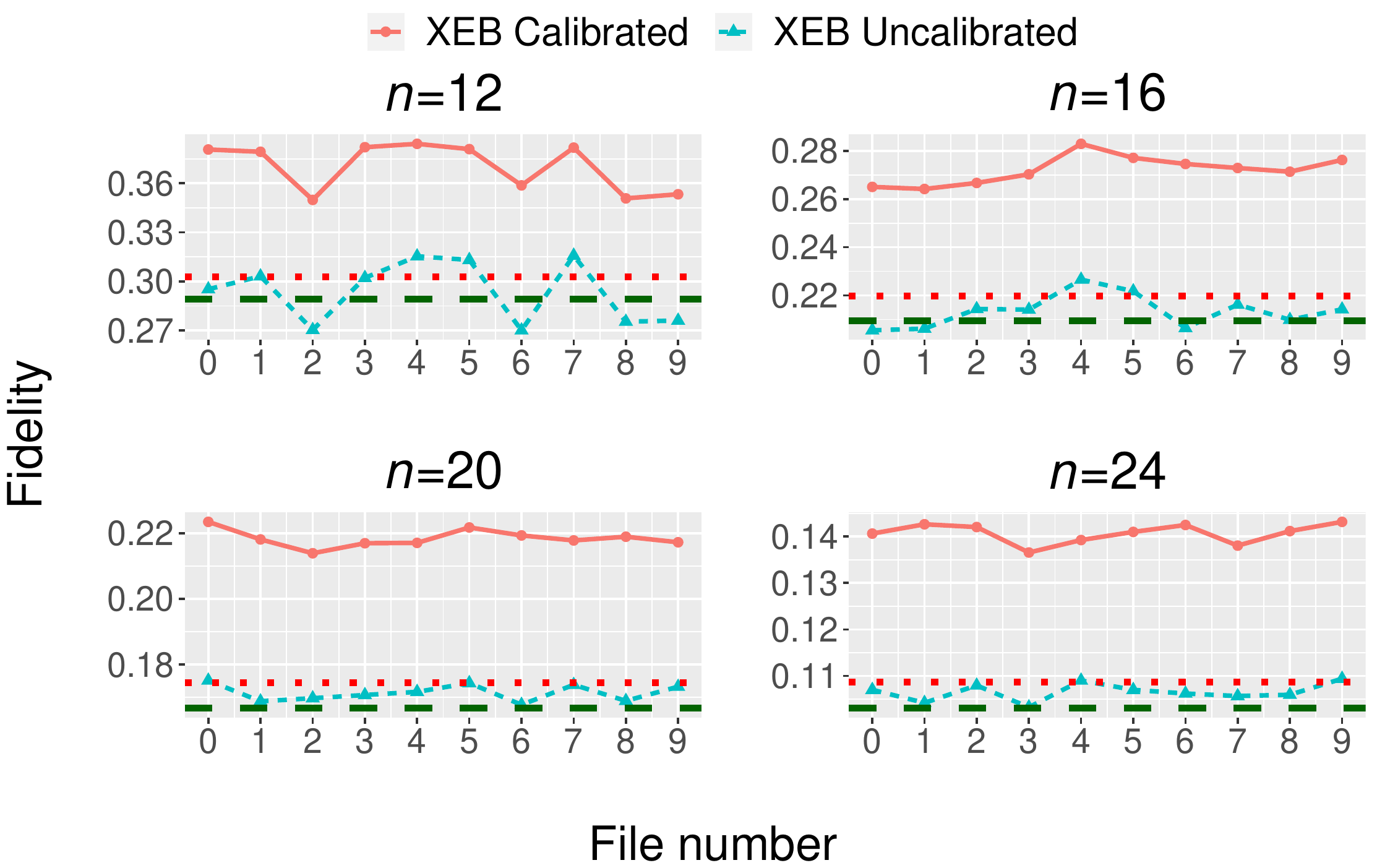}
\caption{The effect of removing the two-gate adjustments (for all two-gates) compared to the ideal effect based on simulations and on Google's estimation program. $\phi$ is taken to be the average $F_{XEB}$. In dotted red is the theoretical reduction, in dark green simulation. 
}
\label{2-gate-calib}
\end{figure}

We also compared the effect of removing all 2-gate adjustments. The first row in Table \ref {T:2gate_adju} describes the ratio (averaged over ten files) between the fidelity of the empirical samples with respect to the circuit where the 2-gate adjustments are removed and the fidelity with respect to the calibrated circuits. The second row describes this ratio based on simulations assuming that the 2-gate adjustments are perfect. The third row describes this ratio based on simulations assuming that the 2-gate adjustments are perfect based on Formulas \eqref{eq:red_fid_cal} and \eqref{eq:red_fid_tot}.  Figure \ref {2-gate-calib} shows the picture for the ten Google files seperately.  


\begin{table}[]
\centering
 \resizebox{\textwidth}{!}{
    \begin{tabular}{|c|c|c|c|c|c|c|c|c|}
    \hline
    $n$&12&14&16&18&20&22&24&26\\
    \hline   
    Empiric &0.793&0.789&0.784&0.782&0.785&0.766&0.757&0.712\\
    Simulation &0.781&0.773&0.770&0.766&0.763&0.744&0.732&0.676\\
    Theory &0.818&0.813&0.808&0.803&0.799&0.783&0.772&0.725\\
    \hline
    \end{tabular}}
    \caption{The effect of the 2-gate adjustments, averaged over ten files: based on the empirical data, based on simulations (assuming the adjustment is perfect), and based on the Formula \eqref{eq:red_fid_tot} (assuming the adjustment is perfect). }
    \label{T:2gate_adju}
\end{table}


{\bf Remarks:} 1. Another property of the calibration that we do not understand is that the effect of 2-gate adjustments for the {\bf EFGH} circuits 
is considerably higher for the 2-gate involving qubits (3,3) and (3,4). 
The effect of removing the adjustments for this 2-gate, that occurs four times in the circuit, is the four big 
troughs
in Figure \ref {calib6}. 

 2. Here is the formula,  provided by the Google team, for the effect of the 2-gate adjustments. 
Assuming that the 2-gate adjustments are  perfect, the  estimate for the decrease in the fidelity due to the removal of the calibration of a single two-gate (whose parameters are $\theta_i$ and $\phi_i$) is given by
    \begin{multline}\label{eq:red_fid_cal}
        \psi_i = \big(4 + \cos(2(\theta_i - \pi/2)) + \cos(\phi_i - \pi/6) \\ 
         + 2*\cos (\theta_i - \pi/2)*(1 + \cos(\phi_i - \pi/6))\big)/10,
    \end{multline}

\noindent
    and the estimate for the overall decrease in the fidelity is estimated by the product of these quantities over all the 2-gates (based on a statistical independence assumption for such different adjustments) is:
    \begin {equation}\label {eq:red_fid_tot}
     R_0 = \prod_{i=1}^{|{\cal G}_2|}\psi_i .
    \end {equation}


\subsection {Summary of Section \ref {s:calib}}
The calibration process accounts for systematic errors for 2-gates and applies certain adjustments to the definition of the circuits. These adjustments are local, namely the adjustments for a 2-gate involving qubits $x$ and $y$ primarily depend on outcomes for 1- and 2-circuits on these qubits. 
Some statistical findings regarding the calibration process are:  (i) The effects of the calibration is large even for a single 2-gate; (ii) there is a large difference between the effect for different 2-gates and even different appearances of the same 2-gate, and; (iii)  the effectiveness of the 2-gate calibrations is remarkable.  

We note that these findings enhance the tension between the calibration process and Google's claim for a ``programmable quantum computer.” The effectiveness of the calibration process 
is especially surprising in view of the local nature of the calibration: mathematically speaking,
we witness a local optimization process reaching a critical point of a
function depending on hundreds of parameters. 

\newpage

\section {Summary} 
In this paper we studied three main issues, mostly from a statistical point of view. 
The first issue, already demonstrated in \cite {RSK22} is the large gap between the Google samples and the Google noise model or any specific noise model we studied. 
This finding weakens the claim made in \cite {Aru+19} of
achieving a successful sampling task, namely achieving approximate sampling
for some specific distribution, and may shed doubt on the potential of NISQ
systems to achieve sampling tasks.

The second issue concerns  Formula (77)  of the Google paper \eqref {e:77} with its
simple independence-like form. This, and the noise and calibration involved in the experimental process, make it surprising that Google’s fidelity estimates appear to be
rather close to Formula (77) for hundreds of different experiments. Surprises
do occur in science, and may lead to important scientific discoveries, but from a statistical point of view, the subsumed independence between  components of systems such as quantum computers, which are known to be sensitive to noise and errors caused by interactions with their environment, is striking.
Related questions arise from the systematic deviation of Formula (77) for patch
circuits, 
and the large difference 
between the ${\cal F}_{XEB}$ fidelity of the two patches.
The remarkable verification of Kalachev et al. \cite {KPY21} and Liu et al. \cite {Liu+22} weakened these concerns by showing that 
Formula (77) provides a good approximation to hundreds of circuits for
which the Google team stated that they had not computed the amplitudes.
USTC’s close 
replications of the Google experiment also alleviate our
concerns regarding Formula (77) and the Google experiment.

The third issue discussed in this paper is 
the calibration process in the Google experiment, the details of which are patented and not fully revealed.  
The facts that circuits must be calibrated and the large effects
of the calibration
adjustments  weaken the claim that Google's experiment describes a programmable quantum computer. 
The close to perfect effectiveness
of the calibration process is surprising especially in-view of its local nature.

We laid the dry facts and findings, and we let the readers make their own interpretation, or rather take note of our concerns and wait for more experimental data from  
future experiments. We hope that our detailed analysis of the Google experiment from \cite {RSK22,KRS22d} and 
this paper will help the community in its assessment of Google's quantum supremacy claims \cite {Aru+19,Aru+19S}, as well as 
other far-reaching claims in the very active field of experimental noisy intermediate-scale quantum computing.

\newpage
\subsection*{ Appendix: Additional information}
\subsubsection *{Distances between various distributions for $n=12,14$}

\begin{table}[H]
\centering
\resizebox{\textwidth}{!}{
    \begin{tabular}{|c|c|c|c|c|c|c|}
    \hline
        \multicolumn{7}{|c|}{\textbf{n=12}} \\ \hline \hline
         Model 1 & Model 2 & \textbf{${\cal X}^2$} & \textbf{$L_1$} & \textbf{$L_2$} & KL &  Cor \\
         \hline
        Google noise model&Google sample&41190&112072&2370&0.040&0.77\\
        &Simulation from Google noise model&4090&35569&707&0.004&0.97\\
        &Simulation from symmetric noise model&5555&41110&813&0.006&0.96\\
        &Simulation from asymmetric noise model&12342&61462&1221&0.012&0.92\\
        &Simulation from a uniform distribution&56297&140502&2983&0.058&$\sim 0$\\
        Google sample&Symmetric noise model&40018&110498&2343&0.039&0.78\\
        &Asymmetric noise model&36786&105468&2252&0.035&0.80\\
        \hline
        \multicolumn{7}{c}{} \\ 
        \multicolumn{7}{c}{} \\ 
        \hline
        \multicolumn{7}{|c|}{\textbf{n=14}} \\ \hline \hline
        Model 1 & Model 2 & \textbf{${\cal X}^2$} & \textbf{$L_1$} & \textbf{$L_2$} & KL &  Cor \\
        \hline 
        Google noise model&Google sample&54286&129436&1339&0.052&0.69\\
        &Simulation from Google noise model&16381&71385&707&0.017&0.88\\
        &Simulation from symmetric noise model&17685&74058&732&0.018&0.87\\
        &Simulation from asymmetric noise model&25347&88504&878&0.025&0.82\\
        &Simulation from a uniform distribution&58825&139950&1473&0.060&$\sim 0$\\
        Google sample&Symmetric noise model&52903&127792&1323&0.052&0.70\\
        &Asymmetric noise model&50093&124112&1290&0.049&0.72\\
        \hline
    \end{tabular}}
    \caption{Distances between the Google sample, the Google noise model, the symmetric noise model, and the asymmetric noise model, on the one side, to different models and simulations based on them, on the other side. For the Google samples, the distances are averaged over the ten Google files. Each simulation is based on 100 repetitions. 
    $\phi$ in the simulations is taken to be the average ${\cal F}_{XEB}$, 0.3701 for $n=12$, and 0.3298 for $n=14$. $\phi_{ro}$, the symmetric readout coefficient, is $\phi/((1-0.038)^n)-\phi$ and the coefficient of the asymmetric readout model is $\phi_{ro}+\phi$, where in both $\phi$ is the average ${\cal F}_{XEB}$.}
    \label{T:distances-full}
\end{table}

\subsubsection*{The ${\cal F}_{XEB}$-fidelities compared to Formula (77) and its simplified versions. 
Values provided in the supplementary data to \cite {Aru+19}.} 
\begin{table}[H]
\centering
 \resizebox{\textwidth}{!}{ \begin{tabular}{||c c c c c c c c c||} 
 \hline
 n & m & circuit type & num 1-gate & num 2-gate  & Formula \eqref {e:77s}  &Formula \eqref{e:77s2} & (77)-Google & Ave ${\cal F}_{XEB}$\\ [0.5ex] 
 \hline\hline
12&14&EFGH&180&60&0.3242&0.3586&0.3862&0.3701\\
14&14&EFGH&210&70&0.2687&0.3023&0.332&0.3298\\
16&14&EFGH&240&80&0.2228&0.2548&0.2828&0.2721\\
18&14&EFGH&270&94&0.1801&0.2069&0.2207&0.2442\\
20&14&EFGH&300&105&0.1483&0.1728&0.1875&0.2184\\
22&14&EFGH&330&116&0.1222&0.1443&0.1554&0.1650\\
24&14&EFGH&360&130&0.0988&0.1171&0.1256&0.1407\\
26&14&EFGH&390&140&0.0819&0.0987&0.1024&0.1141\\
28&14&EFGH&420&150&0.0679&0.0832&0.09072&0.0949\\
30&14&EFGH&450&164&0.0549&0.0676&0.07594&0.0823\\
32&14&EFGH&480&175&0.0452&0.0564&0.06236&0.0713\\
34&14&EFGH&510&186&0.0372&0.0471&0.04592&0.0586\\
36&14&EFGH&540&200&0.0301&0.0383&0.04535&0.0520\\
38&14&EFGH&570&210&0.0249&0.0322&0.03693&$0.0419^{*}$\\
39&14&EFGH&585&217&0.0224&0.0291&0.03073&$0.0343^{*}$\\
40&14&EFGH&600&224&0.0202&0.0262&0.02751&$0.0296^{*}$\\
41&14&EFGH&615&227&0.0186&0.0245&0.02514&$0.0241^{*}$\\
42&14&EFGH&630&234&0.0167&0.0221&0.0229&$0.0229^{*}$\\
43&14&EFGH&645&241&0.015&0.0199&0.02043&$0.0185^{*}$\\
44&14&EFGH&660&248&0.0135&0.0179&0.01901&$0.0184^{*}$\\
45&14&EFGH&675&255&0.0122&0.0161&0.016&$0.0165^{*}$\\
46&14&EFGH&690&259&0.0111&0.0150&0.01485&$0.0154^{*}$\\
47&14&EFGH&705&266&0.0100&0.0135&0.01281&$0.0128^{*}$\\
48&14&EFGH&720&273&0.009&0.0122&0.01134&$0.0121^{*}$\\
49&14&EFGH&735&280&0.0081&0.0109&0.01088&$0.0094^{*}$\\
50&14&EFGH&750&287&0.0073&0.0099&0.009337&$0.0090^{*}$\\
51&14&EFGH&765&294&0.0065&0.0089&0.008978&-\\
53&14&EFGH&795&301&0.0055&0.0077&0.007666&$0.0074^{*}$\\
53&12&ABCDCDAB&689&258&0.0086&0.0115&0.01206&$0.0147^{*}$\\
53&14&ABCDCDAB&795&301&0.0055&0.0077&0.007925&$0.0094^{*}$\\
53&16&ABCDCDAB&901&344&0.0036&0.0052&0.005308&$0.0060^{*}$\\
53&18&ABCDCDAB&1007&387&0.0023&0.0035&0.003555&-\\
53&20&ABCDCDAB&1113&430&0.0015&0.0023&0.002335&$0.0019^{**}$\\
    \hline
\end{tabular}}
\caption{The predictive power of Formulas (77), \eqref {e:77s}, and \eqref {e:77s2} for full circuits.  Based on the supplementary data to \cite {Aru+19}.}
    \label{T:77} 
    \raggedright{
    {\footnotesize * Computed by Kalachev et al. (\cite{KPY21} and private communication).} \\
    {\footnotesize **Computed in \cite{Liu+22} for one circuit.}
    }
\end{table}
\newpage
\subsubsection *{${\cal F}_{XEB}$ for the full, elided and patch circuits, and
Formula (77)}

\begin{table}[H]
\centering
\resizebox{.7\textwidth}{!}{
    \begin{tabular}{|c|c|c|c|c|c|c|}
    \hline
       $n$&Circuit type&m&patch&elided&full&prediction (77)\\
  \hline
12&EFGH&14&0.3604&0.3898&0.3701&0.3862\\
14&EFGH&14&0.3364&0.3282&0.3298&0.3320\\
16&EFGH&14&0.2679&0.2920&0.2721&0.2828\\
18&EFGH&14&0.2520&0.2481&0.2442&0.2207\\
20&EFGH&14&0.2225&0.2144&0.2184&0.1875\\
22&EFGH&14&0.1767&0.1735&0.1650&0.1554\\
24&EFGH&14&0.1394&0.1511&0.1407&0.1256\\
26&EFGH&14&0.1146&0.1157&0.1141&0.1024\\
28&EFGH&14&0.0954&0.09234&0.0949&0.0907\\
30&EFGH&14&0.0807&0.0836&0.0823&0.0759\\
32&EFGH&14&0.0707&0.0706&0.0713&0.0624\\
34&EFGH&14&0.0506&0.0616&0.0586&0.0459\\
36&EFGH&14&0.0501&0.0478&0.0520&0.0454\\
38&EFGH&14&0.0383&0.0440&$0.0419^{*}$&0.0369\\
39&EFGH&14&0.0342&0.0391&$0.0343^{*}$&0.0307\\
40&EFGH&14&0.0278&0.0298&$0.0296^{*}$&0.0275\\
41&EFGH&14&0.0279&0.0294&$0.0241^{*}$&0.0251\\
42&EFGH&14&0.0206&0.0222&$0.0229^{*}$&0.0229\\
43&EFGH&14&0.0194&0.0216&$0.0185^{*}$&0.0204\\
44&EFGH&14&0.0181&0.0197&$0.0184^{*}$&0.0190\\
45&EFGH&14&0.0168&0.0167&$0.0165^{*}$&0.0160\\
46&EFGH&14&0.0150&0.0151&$0.0154^{*}$&0.0149\\
47&EFGH&14&0.0139&0.0148&$0.0128^{*}$&0.0128\\
48&EFGH&14&0.0110&0.0114&$0.0121^{*}$&0.0113\\
49&EFGH&14&0.0104&0.0110&$0.0094^{*}$&0.0109\\
50&EFGH&14&0.0084&0.0088&$0.0090^{*}$&0.0093\\
51&EFGH&14&0.0082&0.0084&-&0.0090\\
53&EFGH&14&0.0072&0.0069&$0.0074^{*}$&0.0077\\
53&ABCDCDAB&12&0.0131&0.0139&$0.0147^{*}$&0.0121\\
53&ABCDCDAB&14&0.0085&0.0090&$0.0094^{*}$&0.0079\\
53&ABCDCDAB&16&0.0054&0.0058&$0.0060^{*}$&0.0053\\
53&ABCDCDAB&18&0.0033&0.0039&-&0.0036\\
53&ABCDCDAB&20&0.0022&0.0022&$0.0019^{**}$&0.0023\\
        \hline
    \end{tabular}}
\caption{Fidelities for the full, elided and patch circuits, compared to Google’s Formula (77), $n=12-53$. The values for patch and elided circuits, the predictions based on Formula (77), and the values for full circuits for $n < 38$, were provided in the supplementary data to \cite {Aru+19}.}
    \label{T:fidelity_all_circuits}
    \raggedright{
    {\footnotesize * Computed by Kalachev et al. (\cite{KPY21} and private communication)} \\
    {\footnotesize **One circuit. Computed in \cite{Liu+22}}
    }
\end{table}
\newpage
\subsubsection *{Data for patch circuits}
The following tables present our computations based on data and programs provided by the Google team.
(There are some small disagreements with the ${\cal F}_{XEB}$-values from Table \ref {T:fidelity_all_circuits}.) 
\begin{table}[H]
\centering
 \resizebox{\textwidth}{!}{
    \begin{tabular}{|c|c|c|c|c|c|c|c|c|c|}
    \hline
        Bitstring & \multicolumn{3}{c|}{First Patch} & \multicolumn{3}{c|}{Second Patch} & \multicolumn{3}{c|}{Combined circuit} \\ \hline \hline 
        ~ & $n$ & \textbf{XEB} & Formula \eqref {e:77s} &   $n$ & \textbf{XEB} & Formula \eqref {e:77s} & Product of \textbf{XEB} & Formula \eqref {e:77s} &  Formula (77)    \\  \hline
12&6&0.5771&0.6022&6&0.5547&0.6022&0.3201&0.3626&0.3862\\
14&6&0.5677&0.6022&8&0.5759&0.4991&0.3269&0.3006&0.3320\\
16&8&0.4928&0.4991&8&0.5506&0.4991&0.2713&0.2491&0.2828\\
18&9&0.4707&0.4488&9&0.5323&0.4488&0.2506&0.2014&0.2207\\
20&9&0.4722&0.4488&11&0.4798&0.3697&0.2266&0.1659&0.1875\\
22&11&0.3825&0.3697&11&0.4635&0.3697&0.1773&0.1367&0.1554\\
24&12&0.3293&0.3324&12&0.4213&0.3324&0.1387&0.1105&0.1256\\
26&14&0.2670&0.2755&12&0.4279&0.3324&0.1142&0.0916&0.1024\\
28&14&0.2695&0.2755&14&0.3509&0.2755&0.0946&0.0759&0.0907\\
30&15&0.2470&0.2477&15&0.3270&0.2477&0.0808&0.0614&0.0759\\
32&17&0.2131&0.2041&15&0.3326&0.2477&0.0709&0.0506&0.0624\\
34&17&0.2068&0.2041&17&0.2437&0.2041&0.0504&0.0417&0.0459\\
36&18&0.1898&0.1835&18&0.2632&0.1835&0.0500&0.0337&0.0454\\
38&19&0.1611&0.1650&19&0.2426&0.1691&0.0391&0.0279&0.0369\\
39&20&0.1432&0.1511&19&0.2436&0.1691&0.0349&0.0256&0.0307\\
40&20&0.1394&0.1511&20&0.2197&0.1521&0.0306&0.0230&0.0275\\
41&21&0.1044&0.1393&20&0.2177&0.1521&0.0227&0.0212&0.0251\\
42&21&0.1149&0.1393&21&0.1930&0.1367&0.0222&0.0190&0.0229\\
43&22&0.0972&0.1253&21&0.1841&0.1367&0.0179&0.0171&0.0204\\
44&22&0.0955&0.1253&22&0.1640&0.1230&0.0157&0.0154&0.0190\\
45&23&0.0879&0.1126&22&0.1643&0.1230&0.0144&0.0138&0.0160\\
46&23&0.1004&0.1126&23&0.1509&0.1126&0.0152&0.0127&0.0149\\
47&24&0.0876&0.1013&23&0.1493&0.1126&0.0131&0.0114&0.0128\\
48&24&0.0931&0.1013&24&0.1347&0.1038&0.0125&0.0105&0.0113\\
49&25&0.0688&0.0911&24&0.1448&0.1038&0.0100&0.0095&0.0109\\
50&25&0.0723&0.0911&25&0.1264&0.0934&0.0091&0.0085&0.0093\\
51&25&0.0717&0.0911&26&0.1170&0.0839&0.0084&0.0076&0.0090\\
        \hline
    \end{tabular}}
    \caption{Fidelities for the patch circuits, circuit type {\bf EFGH}.}
    \label{T:77-patched_combined}
\end{table}

\begin{table}[H]
\centering
 \resizebox{\textwidth}{!}{
    \begin{tabular}{|c|c|c|c|c|c|c|c|c|c|}
    \hline
        $m$ &  \multicolumn{3}{c|}{First Patch} & \multicolumn{3}{c|}{Second Patch} & \multicolumn{3}{c|}{Combined circuit} \\ \hline \hline 
        ~ & $n$ & \textbf{XEB} & Formula \eqref {e:77s} &   $n$ & \textbf{XEB} & Formula \eqref {e:77s} & Product of \textbf {XEB}  & Formula \eqref {e:77s} & Formula (77)  \\  \hline
        12&27&0.0821&0.095&26&0.1563&0.1027&0.0131&0.0098&0.0121\\
14&27&0.0647&0.0769&26&0.1325&0.0839&0.0085&0.0065&0.0079\\
16&27&0.0486&0.0623&26&0.1113&0.0682&0.0054&0.0042&0.0053\\
18&27&0.0354&0.0504&26&0.0855&0.0554&0.0033&0.0028&0.0036\\
20&27&0.0261&0.0409&26&0.0697&0.0453&0.0022&0.0019&0.0023\\
        \hline
    \end{tabular}}
    \caption{Fidelities for the patch circuits, $n=53$, circuit type {\bf ABCDCDAB}.}
    \label{T:77-patched_ABCDCDAB_combined}
\end{table}

\begin{figure}[H]
	\begin{center}
		\includegraphics[width=1.0\textwidth]{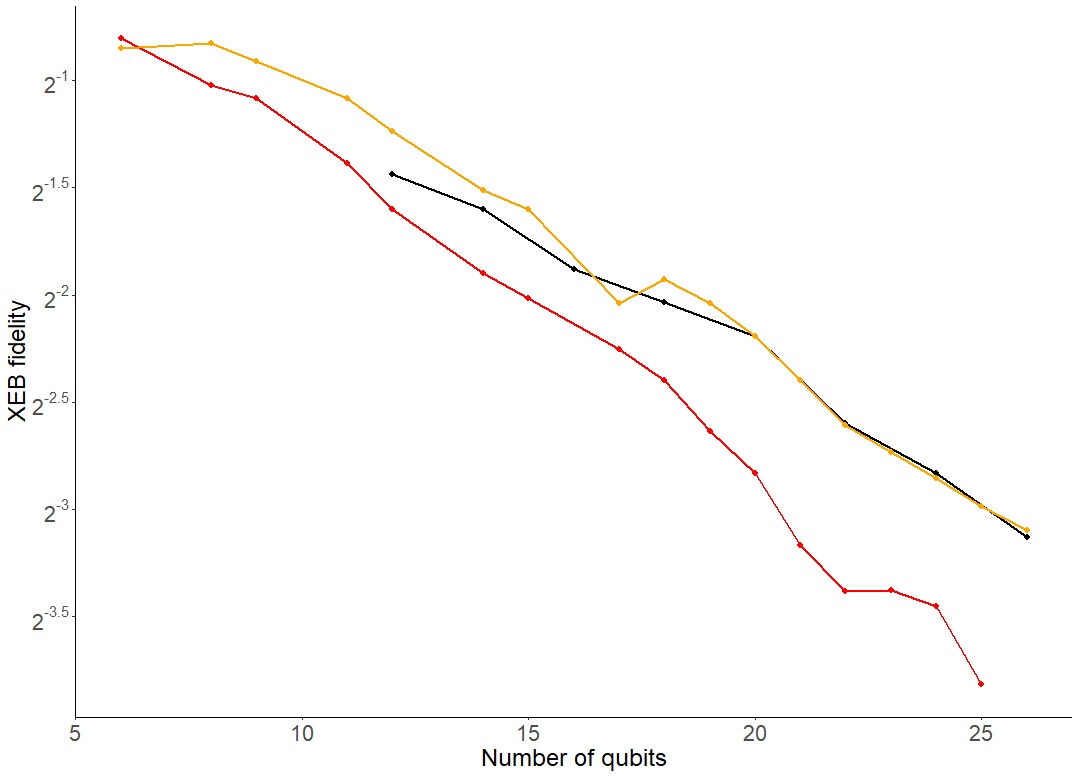}
		\caption{Comparison between average XEB fidelity for the first patch (red) the second patch (orange) and full circuits (circuit type {\bf EFGH}). The behavior for the first patch is notably different. Based on Table \ref {T:77-patched_combined}.} 
		\label{fig:patch++}
	\end{center}
\end{figure}
\bigskip

\medskip
{\small
\noindent
Gil Kalai,  Hebrew University of Jerusalem, Einstein Institute of Mathematics, and\\  Reichman University, Efi Arazi School of Computer Science. \\ {\tt gil.kalai@gmail.com}.

\medskip

\noindent
Yosef Rinott,  Hebrew University of Jerusalem, Federmann Center for the Study of Rationality and Department of Statistics.\\ {\tt yosef.rinott@mail.huji.ac.il}.

\medskip

\noindent
Tomer Shoham,  Hebrew University of Jerusalem, Federmann Center for the Study of Rationality and Department of Computer Science.\\ {\tt tomer.shohamm@gmail.com}.
}

\end {document}